 % EXTERNAL FILES

\input harvmac
\input amssym

% FONTS

% fraktur

\newfam\frakfam
\font\teneufm=eufm10
\font\seveneufm=eufm7
\font\fiveeufm=eufm5
\textfont\frakfam=\teneufm
\scriptfont\frakfam=\seveneufm
\scriptscriptfont\frakfam=\fiveeufm

% black board bold

%\newfam\msbfam
%\font\tenmsb=msbm10
%\font\sevenmsb=msbm7
%\font\fivemsb=msbm5
%\textfont\msbfam=\tenmsb
%\scriptfont\msbfam=\sevenmsb
%\scriptscriptfont\msbfam=\fivemsb
%\def\bb{\fam\msbfam \tenmsb}

% double stroke math

\newfam\dsromfam
\font\tendsrom=dsrom10
\textfont\dsromfam=\tendsrom
\def\ds{\fam\dsromfam \tendsrom}

% bold math italics

\newfam\mbffam
\font\tenmbf=cmmib10
\font\sevenmbf=cmmib7
\font\fivembf=cmmib5
\textfont\mbffam=\tenmbf
\scriptfont\mbffam=\sevenmbf
\scriptscriptfont\mbffam=\fivembf

% bold math cal

\newfam\mbfcalfam
\font\tenmbfcal=cmbsy10
\font\sevenmbfcal=cmbsy7
\font\fivembfcal=cmbsy5
\textfont\mbfcalfam=\tenmbfcal
\scriptfont\mbfcalfam=\sevenmbfcal
\scriptscriptfont\mbfcalfam=\fivembfcal

% math script

\newfam\mscrfam
\font\tenmscr=rsfs10
\font\sevenmscr=rsfs7
\font\fivemscr=rsfs5
\textfont\mscrfam=\tenmscr
\scriptfont\mscrfam=\sevenmscr
\scriptscriptfont\mscrfam=\fivemscr
\def\scr{\fam\mscrfam \tenmscr}

% MACROS

% bras, kets, ...

\def\vev#1{\left\langle #1\right\rangle}

% tilde, hat, bar, ...

\def\tilde{\widetilde}
\def\t{\tilde}
\def\hat{\widehat}

\def\bar{\overline}
\def\b{\bar}
\def\bsq#1{{{\b{#1}}^{\lower 2.5pt\hbox{$\scriptstyle 2$}}}}
\def\bexp#1#2{{{\b{#1}}^{\lower 2.5pt\hbox{$\scriptstyle #2$}}}}
\def\dotexp#1#2{{{#1}^{\lower 2.5pt\hbox{$\scriptstyle #2$}}}}

% basic math

\def\rt2{\sqrt{2}}
\def\half {{1 \over 2}}

\def\d{\partial}

\def\det{\mathop{\rm det}}

\def\Tr{\mathop{\rm Tr}}

% bold greek characters

\font\tenbifull=cmmib10
\font\tenbimed=cmmib7
\font\tenbismall=cmmib5
\textfont9=\tenbifull \scriptfont9=\tenbimed
\scriptscriptfont9=\tenbismall

\mathchardef\bbGamma="7000
\mathchardef\bbDelta="7001
\mathchardef\bbPhi="7002
\mathchardef\bbAlpha="7003
\mathchardef\bbXi="7004
\mathchardef\bbPi="7005
\mathchardef\bbSigma="7006
\mathchardef\bbUpsilon="7007
\mathchardef\bbTheta="7008
\mathchardef\bbPsi="7009
\mathchardef\bbOmega="700A
\mathchardef\bbalpha="710B
\mathchardef\bbbeta="710C
\mathchardef\bbgamma="710D
\mathchardef\bbdelta="710E
\mathchardef\bbepsilon="710F
\mathchardef\bbzeta="7110
\mathchardef\bbeta="7111
\mathchardef\bbtheta="7112
\mathchardef\bbiota="7113
\mathchardef\bbkappa="7114
\mathchardef\bblambda="7115
\mathchardef\bbmu="7116
\mathchardef\bbnu="7117
\mathchardef\bbxi="7118
\mathchardef\bbpi="7119
\mathchardef\bbrho="711A
\mathchardef\bbsigma="711B
\mathchardef\bbtau="711C
\mathchardef\bbupsilon="711D
\mathchardef\bbphi="711E
\mathchardef\bbchi="711F
\mathchardef\bbpsi="7120
\mathchardef\bbomega="7121
\mathchardef\bbvarepsilon="7122
\mathchardef\bbvartheta="7123
\mathchardef\bbvarpi="7124
\mathchardef\bbvarrho="7125
\mathchardef\bbvarsigma="7126
\mathchardef\bbvarphi="7127

% dotted spinor indices

\def\alphadot{{\dot\alpha}}

% bared indices

% bared spinors

\def\sigmabar{\b{\sigma}}
\def\thetabar{\b{\theta}}
\def\thetasq{\theta^2}
\def\thetabarsq{\bsq{\theta}}

% capital cal letters

\def\CC{{\cal C}}

\def\CJ{{\cal J}}

\def\CM{{\cal M}}

\def\CO{{\cal O}}

% double stroke symbols: unit matrix, reals, complex, quaternions, integers, natural numbers

\def\1{{\ds 1}}

% miscellaneous objects

\def\ep{\varepsilon}

\noblackbox

% DRAFTMODE

%\draftmode

% REFERENCES

%\BuicanWS
\lref\BuicanWS{
  M.~Buican, P.~Meade, N.~Seiberg and D.~Shih,
  ``Exploring General Gauge Mediation,''
  JHEP {\bf 0903}, 016 (2009)
  [arXiv:0812.3668 [hep-ph]].
  %%CITATION = JHEPA,0903,016;%%
}

%\MartinZB
\lref\MartinZB{
  S.~P.~Martin,
  ``Generalized messengers of supersymmetry breaking and the sparticle mass
  spectrum,''
  Phys.\ Rev.\  D {\bf 55}, 3177 (1997)
  [arXiv:hep-ph/9608224].
  %%CITATION = PHRVA,D55,3177;%%
}

%\MarquesYU
\lref\MarquesYU{
  D.~Marques,
  ``Generalized messenger sector for gauge mediation of supersymmetry breaking
  and the soft spectrum,''
  JHEP {\bf 0903}, 038 (2009)
  [arXiv:0901.1326 [hep-ph]].
  %%CITATION = JHEPA,0903,038;%%
}

%\tHooftFH
\lref\tHooftFH{
  G.~'t Hooft,
  ``Renormalization Of Massless Yang-Mills Fields,''
  Nucl.\ Phys.\  B {\bf 33}, 173 (1971).
  %%CITATION = NUPHA,B33,173;%%
}

%\MeadeWD
\lref\MeadeWD{
  P.~Meade, N.~Seiberg and D.~Shih,
  ``General Gauge Mediation,''
  Prog.\ Theor.\ Phys.\ Suppl.\  {\bf 177}, 143 (2009)
  [arXiv:0801.3278 [hep-ph]].
  %%CITATION = PTPSA,177,143;%%
}

%\PoppitzXW
\lref\PoppitzXW{
  E.~Poppitz and S.~P.~Trivedi,
  ``Some remarks on gauge-mediated supersymmetry breaking,''
  Phys.\ Lett.\  B {\bf 401}, 38 (1997)
  [arXiv:hep-ph/9703246].
  %%CITATION = PHLTA,B401,38;%%
}

%\GiudiceNI
\lref\GiudiceNI{
  G.~F.~Giudice and R.~Rattazzi,
  ``Extracting Supersymmetry-Breaking Effects from Wave-Function
  Renormalization,''
  Nucl.\ Phys.\  B {\bf 511}, 25 (1998)
  [arXiv:hep-ph/9706540].
  %%CITATION = NUPHA,B511,25;%%
}

%\CheungES
\lref\CheungES{
  C.~Cheung, A.~L.~Fitzpatrick and D.~Shih,
  ``(Extra)Ordinary Gauge Mediation,''
  JHEP {\bf 0807}, 054 (2008)
  [arXiv:0710.3585 [hep-ph]].
  %%CITATION = JHEPA,0807,054;%%
}

%\KomargodskiRZ
\lref\KomargodskiRZ{
  Z.~Komargodski and N.~Seiberg,
  ``From Linear SUSY to Constrained Superfields,''
  JHEP {\bf 0909}, 066 (2009)
  [arXiv:0907.2441 [hep-th]].
  %%CITATION = JHEPA,0909,066;%%
}

%\WeinbergUV
\lref\WeinbergUV{
  S.~Weinberg,
  ``Non-renormalization theorems in non-renormalizable theories,''
  Phys.\ Rev.\ Lett.\  {\bf 80}, 3702 (1998)
  [arXiv:hep-th/9803099].
  %%CITATION = PRLTA,80,3702;%%
}

%\SeibergQJ
\lref\SeibergQJ{
  N.~Seiberg, T.~Volansky and B.~Wecht,
  ``Semi-direct Gauge Mediation,''
  JHEP {\bf 0811}, 004 (2008)
  [arXiv:0809.4437 [hep-ph]].
  %%CITATION = JHEPA,0811,004;%%
}

%\CohenQC
\lref\CohenQC{
  A.~G.~Cohen, T.~S.~Roy and M.~Schmaltz,
  ``Hidden sector renormalization of MSSM scalar masses,''
  JHEP {\bf 0702}, 027 (2007)
  [arXiv:hep-ph/0612100].
  %%CITATION = JHEPA,0702,027;%%
}

%\RoyNZ
\lref\RoyNZ{
  T.~S.~Roy and M.~Schmaltz,
  ``A hidden solution to the $mu/B_mu$ problem in gauge mediation,''
  Phys.\ Rev.\  D {\bf 77}, 095008 (2008)
  [arXiv:0708.3593 [hep-ph]].
  %%CITATION = PHRVA,D77,095008;%%
}

%\ArgurioGE
\lref\ArgurioGE{
  R.~Argurio, M.~Bertolini, G.~Ferretti and A.~Mariotti,
  ``Patterns of Soft Masses from General Semi-Direct Gauge Mediation,''
  arXiv:0912.0743 [hep-ph].
  %%CITATION = ARXIV:0912.0743;%%
}

%\ElvangGK
\lref\ElvangGK{
  H.~Elvang and B.~Wecht,
  ``Semi-Direct Gauge Mediation with the 4-1 Model,''
  JHEP {\bf 0906}, 026 (2009)
  [arXiv:0904.4431 [hep-ph]].
  %%CITATION = JHEPA,0906,026;%%
}

%\IntriligatorFR
\lref\IntriligatorFR{
  K.~A.~Intriligator and M.~Sudano,
  ``Comments on General Gauge Mediation,''
  JHEP {\bf 0811}, 008 (2008)
  [arXiv:0807.3942 [hep-ph]].
  %%CITATION = JHEPA,0811,008;%%
}

%\DistlerBT
\lref\DistlerBT{
  J.~Distler and D.~Robbins,
  ``General F-Term Gauge Mediation,''
  arXiv:0807.2006 [hep-ph].
  %%CITATION = ARXIV:0807.2006;%%
}

%\ArkaniHamedFB
\lref\ArkaniHamedFB{
  N.~Arkani-Hamed and S.~Dimopoulos,
  ``Supersymmetric unification without low energy supersymmetry and  signatures
  for fine-tuning at the LHC,''
  JHEP {\bf 0506}, 073 (2005)
  [arXiv:hep-th/0405159].
  %%CITATION = JHEPA,0506,073;%%
}

%\GiudiceTC
\lref\GiudiceTC{
  G.~F.~Giudice and A.~Romanino,
  ``Split supersymmetry,''
  Nucl.\ Phys.\  B {\bf 699}, 65 (2004)
  [Erratum-ibid.\  B {\bf 706}, 65 (2005)]
  [arXiv:hep-ph/0406088].
  %%CITATION = NUPHA,B699,65;%%
}

%\KaplanAC
\lref\KaplanAC{
  D.~E.~Kaplan, G.~D.~Kribs and M.~Schmaltz,
  ``Supersymmetry breaking through transparent extra dimensions,''
  Phys.\ Rev.\  D {\bf 62}, 035010 (2000)
  [arXiv:hep-ph/9911293].
  %%CITATION = PHRVA,D62,035010;%%
}

%\ChackoMI
\lref\ChackoMI{
  Z.~Chacko, M.~A.~Luty, A.~E.~Nelson and E.~Ponton,
  ``Gaugino mediated supersymmetry breaking,''
  JHEP {\bf 0001}, 003 (2000)
  [arXiv:hep-ph/9911323].
  %%CITATION = JHEPA,0001,003;%%
}

%\ArkaniHamedKJ
\lref\ArkaniHamedKJ{
  N.~Arkani-Hamed, G.~F.~Giudice, M.~A.~Luty and R.~Rattazzi,
  ``Supersymmetry-breaking loops from analytic continuation into  superspace,''
  Phys.\ Rev.\  D {\bf 58}, 115005 (1998)
  [arXiv:hep-ph/9803290].
  %%CITATION = PHRVA,D58,115005;%%
}

%\GiudiceBP
\lref\GiudiceBP{
  G.~F.~Giudice and R.~Rattazzi,
  ``Theories with gauge-mediated supersymmetry breaking,''
  Phys.\ Rept.\  {\bf 322}, 419 (1999)
  [arXiv:hep-ph/9801271].
  %%CITATION = PRPLC,322,419;%%
}

%\FerraraPZ
\lref\FerraraPZ{
  S.~Ferrara and B.~Zumino,
  ``Transformation Properties Of The Supercurrent,''
  Nucl.\ Phys.\  B {\bf 87}, 207 (1975).
  %%CITATION = NUPHA,B87,207;%%
}

%\AffleckXZ
\lref\AffleckXZ{
  I.~Affleck, M.~Dine and N.~Seiberg,
  ``Dynamical Supersymmetry Breaking In Four-Dimensions And Its
  Phenomenological Implications,''
  Nucl.\ Phys.\  B {\bf 256}, 557 (1985).
  %%CITATION = NUPHA,B256,557;%%
}

%\IbeWP
\lref\IbeWP{
  M.~Ibe, Y.~Nakayama and T.~T.~Yanagida,
  ``Conformal gauge mediation,''
  Phys.\ Lett.\  B {\bf 649}, 292 (2007)
  [arXiv:hep-ph/0703110].
  %%CITATION = PHLTA,B649,292;%%
}

%\RandallZI
\lref\RandallZI{
  L.~Randall,
  ``New mechanisms of gauge mediated supersymmetry breaking,''
  Nucl.\ Phys.\  B {\bf 495}, 37 (1997)
  [arXiv:hep-ph/9612426].
  %%CITATION = NUPHA,B495,37;%%
}

%\DimopoulosIG
\lref\DimopoulosIG{
  S.~Dimopoulos and G.~F.~Giudice,
  ``Multi-messenger theories of gauge-mediated supersymmetry breaking,''
  Phys.\ Lett.\  B {\bf 393}, 72 (1997)
  [arXiv:hep-ph/9609344].
  %%CITATION = PHLTA,B393,72;%%
}

%\NakayamaCF
\lref\NakayamaCF{
  Y.~Nakayama, M.~Taki, T.~Watari and T.~T.~Yanagida,
  ``Gauge mediation with D-term SUSY breaking,''
  Phys.\ Lett.\  B {\bf 655}, 58 (2007)
  [arXiv:0705.0865 [hep-ph]].
  %%CITATION = PHLTA,B655,58;%%
}

%\PoppitzTX
\lref\PoppitzTX{
  E.~Poppitz and L.~Randall,
  ``Low-energy Kahler potentials in supersymmetric gauge theories with (almost)
  flat directions,''
  Phys.\ Lett.\  B {\bf 336}, 402 (1994)
  [arXiv:hep-th/9407185].
  %%CITATION = PHLTA,B336,402;%%
}

%\BaggerHH
\lref\BaggerHH{
  J.~Bagger, E.~Poppitz and L.~Randall,
  ``The R axion from dynamical supersymmetry breaking,''
  Nucl.\ Phys.\  B {\bf 426}, 3 (1994)
  [arXiv:hep-ph/9405345].
  %%CITATION = NUPHA,B426,3;%%
}

%\DineYW
\lref\DineYW{
  M.~Dine and A.~E.~Nelson,
  ``Dynamical supersymmetry breaking at low-energies,''
  Phys.\ Rev.\  D {\bf 48}, 1277 (1993)
  [arXiv:hep-ph/9303230].
  %%CITATION = PHRVA,D48,1277;%%
}

%\DineVC
\lref\DineVC{
  M.~Dine, A.~E.~Nelson and Y.~Shirman,
  ``Low-Energy Dynamical Supersymmetry Breaking Simplified,''
  Phys.\ Rev.\  D {\bf 51}, 1362 (1995)
  [arXiv:hep-ph/9408384].
  %%CITATION = PHRVA,D51,1362;%%
}

%\DineAG
\lref\DineAG{
  M.~Dine, A.~E.~Nelson, Y.~Nir and Y.~Shirman,
  ``New tools for low-energy dynamical supersymmetry breaking,''
  Phys.\ Rev.\  D {\bf 53}, 2658 (1996)
  [arXiv:hep-ph/9507378].
  %%CITATION = PHRVA,D53,2658;%%
}

%\KomargodskiAX
\lref\KomargodskiAX{
  Z.~Komargodski and N.~Seiberg,
  ``mu and General Gauge Mediation,''
  JHEP {\bf 0903}, 072 (2009)
  [arXiv:0812.3900 [hep-ph]].
  %%CITATION = JHEPA,0903,072;%%
}

%\CarpenterWI
\lref\CarpenterWI{
  L.~M.~Carpenter, M.~Dine, G.~Festuccia and J.~D.~Mason,
  ``Implementing General Gauge Mediation,''
  Phys.\ Rev.\  D {\bf 79}, 035002 (2009)
  [arXiv:0805.2944 [hep-ph]].
  %%CITATION = PHRVA,D79,035002;%%
}

%\IntriligatorBE
\lref\IntriligatorBE{
  K.~Intriligator and M.~Sudano,
  ``General Gauge Mediation with Gauge Messengers,''
  arXiv:1001.5443 [hep-ph].
  %%CITATION = ARXIV:1001.5443;%%
}

%\DineZA
\lref\DineZA{
  M.~Dine, W.~Fischler and M.~Srednicki,
  ``Supersymmetric Technicolor,''
  Nucl.\ Phys.\  B {\bf 189}, 575 (1981).
  %%CITATION = NUPHA,B189,575;%%
}

%\DimopoulosAU
\lref\DimopoulosAU{
  S.~Dimopoulos and S.~Raby,
  ``Supercolor,''
  Nucl.\ Phys.\  B {\bf 192}, 353 (1981).
  %%CITATION = NUPHA,B192,353;%%
}

%\DineGU
\lref\DineGU{
  M.~Dine and W.~Fischler,
  ``A Phenomenological Model Of Particle Physics Based On Supersymmetry,''
  Phys.\ Lett.\  B {\bf 110}, 227 (1982).
  %%CITATION = PHLTA,B110,227;%%
}

%\NappiHM
\lref\NappiHM{
  C.~R.~Nappi and B.~A.~Ovrut,
  ``Supersymmetric Extension Of The SU(3) X SU(2) X U(1) Model,''
  Phys.\ Lett.\  B {\bf 113}, 175 (1982).
  %%CITATION = PHLTA,B113,175;%%
}

%\AlvarezGaumeWY
\lref\AlvarezGaumeWY{
  L.~Alvarez-Gaume, M.~Claudson and M.~B.~Wise,
  ``Low-Energy Supersymmetry,''
  Nucl.\ Phys.\  B {\bf 207}, 96 (1982).
  %%CITATION = NUPHA,B207,96;%%
}

%\DimopoulosGM
\lref\DimopoulosGM{
  S.~Dimopoulos and S.~Raby,
  ``Geometric Hierarchy,''
  Nucl.\ Phys.\  B {\bf 219}, 479 (1983).
  %%CITATION = NUPHA,B219,479;%%
}

%\DineYW
\lref\DineYW{
  M.~Dine and A.~E.~Nelson,
  ``Dynamical supersymmetry breaking at low-energies,''
  Phys.\ Rev.\  D {\bf 48}, 1277 (1993)
  [arXiv:hep-ph/9303230].
  %%CITATION = PHRVA,D48,1277;%%
}

%\DineVC
\lref\DineVC{
  M.~Dine, A.~E.~Nelson and Y.~Shirman,
  ``Low-Energy Dynamical Supersymmetry Breaking Simplified,''
  Phys.\ Rev.\  D {\bf 51}, 1362 (1995)
  [arXiv:hep-ph/9408384].
  %%CITATION = PHRVA,D51,1362;%%
}

%\DineAG
\lref\DineAG{
  M.~Dine, A.~E.~Nelson, Y.~Nir and Y.~Shirman,
  ``New tools for low-energy dynamical supersymmetry breaking,''
  Phys.\ Rev.\  D {\bf 53}, 2658 (1996)
  [arXiv:hep-ph/9507378].
  %%CITATION = PHRVA,D53,2658;%%
}

%\BuicanVV
\lref\BuicanVV{
  M.~Buican and Z.~Komargodski,
  ``Soft Terms from Broken Symmetries,''
  arXiv:0909.4824 [hep-ph].
  %%CITATION = ARXIV:0909.4824;%%
}

%\IntriligatorBE
\lref\IntriligatorBE{
  K.~Intriligator and M.~Sudano,
  ``General Gauge Mediation with Gauge Messengers,''
  arXiv:1001.5443 [hep-ph].
  %%CITATION = ARXIV:1001.5443;%%
}

%\RandallZI
\lref\RandallZI{
  L.~Randall,
  ``New mechanisms of gauge mediated supersymmetry breaking,''
  Nucl.\ Phys.\  B {\bf 495}, 37 (1997)
  [arXiv:hep-ph/9612426].
  %%CITATION = NUPHA,B495,37;%%
}

%\PerezNG
\lref\PerezNG{
  G.~Perez, T.~S.~Roy and M.~Schmaltz,
  ``Phenomenology of SUSY with scalar sequestering,''
  Phys.\ Rev.\  D {\bf 79}, 095016 (2009)
  [arXiv:0811.3206 [hep-ph]].
  %%CITATION = PHRVA,D79,095016;%%
}

%\SeibergVC
\lref\SeibergVC{
  N.~Seiberg,
  ``Naturalness Versus Supersymmetric Non-renormalization Theorems,''
  Phys.\ Lett.\  B {\bf 318}, 469 (1993)
  [arXiv:hep-ph/9309335].
  %%CITATION = PHLTA,B318,469;%%
}

%\KomargodskiJF
\lref\KomargodskiJF{
  Z.~Komargodski and D.~Shih,
  ``Notes on SUSY and R-Symmetry Breaking in Wess-Zumino Models,''
  JHEP {\bf 0904}, 093 (2009)
  [arXiv:0902.0030 [hep-th]].
  %%CITATION = JHEPA,0904,093;%%
}

%\DineME
\lref\DineME{
  M.~Dine and N.~Seiberg,
  ``Comments on Quantum Effects in Supergravity Theories,''
  JHEP {\bf 0703}, 040 (2007)
  [arXiv:hep-th/0701023].
  %%CITATION = JHEPA,0703,040;%%
}

%\ShadmiMD
\lref\ShadmiMD{
  Y.~Shadmi,
  ``Gauge-mediated supersymmetry breaking without fundamental singlets,''
  Phys.\ Lett.\  B {\bf 405}, 99 (1997)
  [arXiv:hep-ph/9703312].
  %%CITATION = PHLTA,B405,99;%%
}

%\IntriligatorDD
\lref\IntriligatorDD{
  K.~A.~Intriligator, N.~Seiberg and D.~Shih,
  ``Dynamical SUSY breaking in meta-stable vacua,''
  JHEP {\bf 0604}, 021 (2006)
  [arXiv:hep-th/0602239].
  %%CITATION = JHEPA,0604,021;%%
}

%\KomargodskiRB
\lref\KomargodskiRB{
  Z.~Komargodski and N.~Seiberg,
  ``Comments on Supercurrent Multiplets, Supersymmetric Field Theories and
  Supergravity,''
  arXiv:1002.2228 [hep-th].
  %%CITATION = ARXIV:1002.2228;%%
}

%\KuzenkoAM
\lref\KuzenkoAM{
  S.~M.~Kuzenko,
  ``Variant supercurrent multiplets,''
  arXiv:1002.4932 [hep-th].
  %%CITATION = ARXIV:1002.4932;%%
}

% TITLE

\rightline{PUPT-2332}
\Title{\vbox{\baselineskip12pt \hbox{}}} {\vbox{\centerline{General Messenger Gauge Mediation}}}
\smallskip
\centerline{Thomas T. Dumitrescu,$^1$ Zohar Komargodski,$^2$ Nathan Seiberg$^2$ and David Shih$^{2,3}$}
\smallskip
\bigskip
\centerline{$^1${\it Department of Physics, Princeton University, Princeton, NJ 08544, USA}}
\centerline{$^2${\it  School of Natural Sciences, Institute for Advanced Study, Princeton, NJ 08540, USA}}
\centerline{$^3${\it  Department of Physics and Astronomy, Rutgers University, Piscataway, NJ 08854, USA}}
\bigskip
\vskip 1cm

\noindent We discuss theories of gauge mediation in which the hidden sector
consists of two subsectors which are weakly coupled to each other. One sector is
made up of messengers and the other breaks supersymmetry. Each sector by
itself may
be strongly coupled. We provide a unifying framework for such theories and
discuss their predictions in different settings. We show how this framework
incorporates all known models of messengers.
In the case of weakly-coupled messengers interacting with spurions through the superpotential, we prove that the
sfermion mass-squared is positive, and furthermore, that there is a lower
bound on the ratio of the sfermion mass to the gaugino mass.

\bigskip

\Date{March 2010}

\newsec{Introduction}

Supersymmetry at the TeV scale is a leading candidate for physics
beyond the Standard Model (SM). If it is realized in nature, it must be
spontaneously broken in a hidden sector, and this breaking must then be
mediated to the Supersymmetric Standard Model~(SSM). In general, the mediation
is highly constrained by precise experimental tests of flavor symmetry in the
SM. Gauge
mediation~\refs{\DineZA\DimopoulosAU\DineGU\NappiHM\AlvarezGaumeWY \DimopoulosGM
\DineYW\DineVC-\DineAG}\ solves this SUSY
flavor problem by postulating that the hidden sector and the SSM only communicate via
the SM gauge interactions. Since the
gauge interactions are flavor blind, they automatically result in
a flavor-universal SSM spectrum consistent with experiment.

Many different gauge mediation models have been constructed,
giving rise to a wide variety of predictions. In~\MeadeWD, General Gauge Mediation (GGM) was formulated in order to
incorporate these models into a uniform framework. The GGM setup consists of two sectors: a SUSY-breaking hidden sector with a weakly-gauged global symmetry that includes the SM gauge group, and a visible sector that includes the SSM. The defining assumption of GGM is that these two sectors completely decouple when the SM gauge couplings are taken to zero.\foot{Given this definition, GGM does not address the $\mu$-problem of gauge mediation; it also does not allow for gauge messengers. See~\KomargodskiAX\ and~\refs{\BuicanVV,\IntriligatorBE} for extensions of GGM
in these directions.}
By treating the SM gauge interactions perturbatively, but allowing for complicated (in principle strongly-coupled) hidden-sector dynamics, it was possible to
deduce the most general, model-independent features of gauge mediation. In particular, it was shown that
the parameter space of GGM consists of three complex gaugino masses,
and three real parameters which determine the sfermion masses.

In this paper we will focus on a subset of the models described by
GGM. Many of the known gauge mediation models contain hidden-sector fields which are charged under the SM gauge group and
acquire a non-supersymmetric spectrum, but do not themselves
participate in the SUSY-breaking dynamics; such fields are called
messengers. The prime example of this scenario is Minimal Gauge
Mediation (MGM)~\refs{\DineYW\DineVC-\DineAG}, which serves as the foundation
for many
phenomenological studies of gauge mediation. In MGM, a pair of messengers is
weakly coupled to a hidden-sector singlet
$X$ (a spurion) via the superpotential. SUSY-breaking is
parameterized by the $F$-term vev of $X$, which is assumed to
result from the dynamics of the hidden sector. The minimal setup
has been extended by giving the messengers arbitrary supersymmetric masses and
allowing the spurion to interact with the messengers through arbitrary Yukawa couplings~\CheungES. Models
of this type were used in~\CarpenterWI\ to attain the correct
number of GGM parameters, but without covering the full parameter
space. The full parameter space was covered in~\BuicanWS\ by also giving the messengers diagonal SUSY-breaking masses through a $D$-term spurion. This shows that the entire GGM parameter space is theoretically
accessible, even just with models of messengers.

Our goal in this paper is to study general theories with a messenger sector. The motivation for GGM was to elucidate the most general predictions
common to all gauge mediation models; here we would like to understand the most
general consequences of having a messenger sector.
The hidden sector of GGM now itself consists of two subsectors: a
sector of SM-singlets in which SUSY is spontaneously broken at the
scale~$\sqrt{F}$ (this sector is denoted by subscript~$h$), and a supersymmetric
messenger sector characterized by the scale $M$ (this sector is denoted by subscript~$m$),
whose global symmetry contains the SM gauge group. We allow the
most general interactions between the SUSY-breaking sector and the messenger
sector, but we assume that these interactions are weak so that
perturbation theory is applicable:
\eqn\intlag{\delta {\scr L} =
{16\t \lambda \over \Lambda^{\t \Delta_h + \t \Delta_m - 2}} \int
d^4 \theta \, \t \CO_h \t \CO_m + {4\lambda \over \Lambda^{\Delta_h
+ \Delta_m -3}} \int d^2 \theta \, \CO_h \CO_m + {\rm c.c.}~.}
We will refer to the two terms in~\intlag\ and their respective complex conjugates as K\"ahler potential and superpotential interactions. In general, the SUSY-breaking sector and the messenger sector are described by effective theories valid below some UV scale $\Lambda \gg M, \sqrt{F}$. The operators $\CO_h, \t \CO_h$ and $\CO_m, \t \CO_m$ belong to these sectors, respectively. $\CO_h, \CO_m$ are chiral superfields of dimensions $\Delta_h, \Delta_m$, while $\t \CO_h, \t \CO_m$ are unconstrained (in general complex) superfields of dimensions $\t \Delta_h, \t \Delta_m$. Finally, $\lambda, \t \lambda$ are dimensionless couplings; the numerical factors are for later convenience. We refer to this setup as General Messenger Gauge Mediation (GMGM).

Since GMGM is based on an effective Lagrangian valid below the scale $\Lambda$, we must clarify the UV-sensitivity of this setup.   If the interactions \intlag\ are renormalizable, their contributions to the soft masses are $\Lambda$-independent and can always be trusted.  In general, there may be additional non-renormalizable interactions at the scale $\Lambda$ which could contribute to the soft masses. These operators are necessarily $\Lambda$-suppressed, and their contribution to the soft masses is subdominant.  The situation is more subtle if the interactions \intlag\ are not renormalizable.
Now we can only trust our calculations if we know all contributing operators at the
scale~$\Lambda$, unless we can argue that the effects of certain such operators are subdominant. Throughout this paper, we will assume that one of these conditions is satisfied.
Note that in some cases, we are {\it forced} to include additional operators at the scale $\Lambda$, which act as counterterms for divergences in the effective theory.  Logarithmic divergences are an exception, since we can always trust the coefficient of the leading logarithm.

As we will see, the GMGM framework includes the aforementioned weakly-coupled spurion models with general $F$- and $D$-term
SUSY-breaking parameters. It  also includes models with hidden-sector gauge dynamics, such as those described
in~\refs{\RandallZI,\ArgurioGE},
the Semi-Direct Gauge Mediation models studied in~\refs{\IbeWP\SeibergQJ-\ElvangGK}, and messenger models with strong dynamics in the hidden
sector~\refs{\CohenQC\RoyNZ-\PerezNG}. In addition, the framework in principle makes it possible to study models in which the messengers themselves are strongly coupled.

The foundation for our calculations in GMGM is provided by the simple formulas
derived in~\BuicanWS\ for the gaugino mass and the sfermion mass-squared in
GGM:
\eqn\ggmsoftmasses{\eqalign{
 & M_{\t g} = {g^2\over 4}  \int d^4x\,B(x)~,\cr
 & m^2_{\t f} =-{g^4Y^2\over128\pi^2}\int d^4x\,A(x)\log (x^2M^2) ~,
 }}
where $B(x)$ and $A(x)$ are hidden-sector correlation functions defined as
follows:
\eqn\ggmfnsdef{\eqalign{& B(x) = \vev{Q^2(J(x)J(0))}~,\cr
& A(x) = \vev{Q^4(J(x)J(0))}~.}}
For
simplicity, we take the SM gauge group to be $U(1)$ throughout this
paper. The group theory factors for the full case $SU(3) \times
SU(2) \times U(1)$ can be easily restored. As described in~\refs{\MeadeWD,
\BuicanWS}, $J(x)$ is the bottom component of the $U(1)$ current
superfield through which the messengers couple to the SM gauge field, $g$ is
the gauge coupling, and~$Y$ is the charge of the sfermion.\foot{The
GGM formalism is briefly reviewed in appendix A, where we also collect some new
results on the GGM correlation functions. Note that our definition of $B(x)$ in~\ggmfnsdef\ differs from the definition in~\refs{\MeadeWD,
\BuicanWS} by a factor of~$4$.} From~\ggmsoftmasses, we see that
the problem of determining the visible-sector soft masses reduces to calculating
the correlators $B(x)$ and $A(x)$.

In section~2, we give a more detailed definition of GMGM. We expand in powers of the interactions~\intlag\ to obtain leading-order
formulas for the $B$- and $A$-correlators. The formulas simplify because they
factorize into products of correlators which are evaluated
separately in the SUSY-breaking sector and the messenger sector. At
leading order, the SUSY-breaking correlators are relatively simple, while the messenger correlators are always supersymmetric. As a
general consequence of the formalism, we will show that the gaugino mass coming
from K\"ahler potential interactions is typically suppressed relative to the corresponding sfermion mass. In particular, we will show that certain leading-order K\"ahler potential
contributions to the gaugino
mass vanish identically.

The GMGM formalism also leads to a controlled expansion in ${F \over M^2}\ll 1$ when the SUSY-breaking splittings in the messenger sector are small. In this
limit we recover the qualitative
behavior of the soft masses in many known models. A different limit leads to
the well-known spurion regime in which models of messengers are commonly
studied.

Section~3  explains
how some of the GMGM formulas of section~2 can be rewritten in terms of
supersymmetric deformations of the messenger-sector Lagrangian. Using this
technique, we show more conceptually why certain K\"ahler potential
interactions do not generate gaugino masses at leading order.

In section~4, we apply the GMGM formalism to the much-studied case
of weakly-coupled, renormalizable spurion models and calculate the leading-order
soft masses in these models.

For the case of messengers coupling to spurions
through a general superpotential, formulas for the soft masses were first obtained in~\CheungES\ using the technique of wavefunction
renormalization \GiudiceNI. As we will show, these formulas imply that the sfermion
mass-squared is always positive, and moreover that the ratio of
the sfermion mass-squared to the gaugino mass-squared is bounded from below. For
example, with $N$ messengers and SM gauge
group $U(1)$, this ratio satisfies:
 \eqn\MGMineq{{m^2_{\t f} \over M^2_{\t g}} \geq
{Y^2\over N}~.} This inequality explains why models in which spurions only couple through the superpotential cannot cover the
parameter space of GGM \CarpenterWI.

When the spurions couple to the messengers through K\"ahler potential
interactions, the GMGM formulas imply that the leading-order gaugino
mass always vanishes. In some cases, we reinterpret this vanishing as a
consequence of the rescaling
anomaly. We also use our results for
spurion models to discuss a particular limit of Semi-Direct Gauge Mediation. In
the regime we study, we find that without considerable fine-tuning, the sfermion
mass is always much greater than the gaugino mass.

In appendix B, we reanalyze weakly-coupled spurion models in more detail. We use the
techniques of section~3 to rederive the leading-order soft masses. By directly applying the formulas from \refs{\BuicanWS}, we also derive simple expressions for the full, all-orders soft masses. These were first obtained in \MarquesYU, and we find complete agreement. Finally, we discuss the limitations of the wavefunction renormalization technique  (even for small SUSY-breaking), and we explain why it happens to give correct
answers for the soft masses in weakly-coupled spurion models.

\newsec{General Messenger Gauge Mediation}

In this section we give a more detailed definition of General Messenger Gauge
Mediation (GMGM), and we show how the $B$- and $A$-correlators in \ggmfnsdef\
simplify in this framework. A general consequence of the formalism is that
certain K\"ahler potential interactions do not generate a gaugino mass at
leading order. Finally, we discuss two simplifying limits of the GMGM framework:
the limit of small SUSY-breaking, and the spurion limit.

\subsec{Definition of the Framework}

Our definition of GMGM consists of the following three sectors:
\vskip5pt

\item{1.)} A visible sector consisting of the SSM with gauge group $G_{SM}=SU(3)\times SU(2)\times U(1)$ and characteristic scale $M_{\rm weak} \sim 100 \, {\rm GeV}$.

\item{2.)} A messenger sector (denoted by subscript $m$) whose global symmetry group contains~$G_{SM}$. All mass scales in this sector are of order $M$; there are no massless particles. The messenger sector may be strongly coupled.

\item{3.)} A SUSY-breaking sector (denoted by subscript $h$) consisting of
$G_{SM}$-singlets. For simplicity we assume that the scale of all masses and
correlation functions is set by the strength $\sqrt F$ of SUSY-breaking ($F^2$
is the total vacuum energy-density). Like the messenger sector, the
SUSY-breaking sector may be strongly coupled.

\vskip5pt

\noindent Note that the messenger sector (2) and the SUSY-breaking sector (3)
of GMGM together make up what is called the hidden sector in GGM. The visible
sector and the messenger sector interact through the visible-sector gauge
fields; they decouple when the visible-sector gauge couplings vanish. Since the
SUSY-breaking sector is neutral under $G_{SM}$, the current multiplet
which enters the GGM correlators in~\ggmsoftmasses\ and~\ggmfnsdef\ only contains
messenger-sector fields.

A key element of the GMGM framework is the assumption that the interactions
between the SUSY-breaking sector and the messenger sector are ${\it weak}$ and
can be treated in perturbation theory. This is what will allow us to simplify
the $B$- and $A$-correlators in~\ggmfnsdef. With this assumption, the
interactions between the SUSY-breaking sector and the messenger sector take the
general form \intlag, which we repeat here for convenience:
\eqn\genint{\delta {\scr L} = {16\t \lambda \over \Lambda^{\t \Delta_h + \t \Delta_m - 2}} \int d^4 \theta \, \t \CO_h \t \CO_m + {4\lambda \over \Lambda^{\Delta_h + \Delta_m - 3}} \int d^2 \theta \, \CO_h \CO_m + \, {\rm c.c.}~.}
The quantities appearing in~\genint\ were defined below~\intlag.
We refer to the two terms in~\genint\ together with their respective complex
conjugates as K\"ahler potential and superpotential interactions.
In general, the interaction Lagrangian~$\delta {\scr L}$ might contain several
such terms.

\subsec{Soft Masses in GMGM}

It is straightforward to expand in powers of the interactions \genint\ and
compute the leading-order contributions to the $B$- and $A$-correlators in
\ggmfnsdef. We will organize the presentation in
terms of the different types of operators that can couple the
SUSY-breaking sector and the messenger sector. In each case, we will see how the $B$- and $A$-correlators factorize into
separate correlators evaluated in these two sectors, and we will discuss the implications for the gaugino and sfermion masses.

\medskip\medskip\medskip

\centerline{\it Superpotential Interactions}

\eqn\spotlint{\eqalign{
\delta{\scr L} & =  {4 \lambda \over \Lambda^{\Delta_h + \Delta_m - 3}} \int d^2
\theta \, \CO_h \CO_m + \, {\rm c.c.}\cr
 & = {\lambda \over  \Lambda^{\Delta_h +
\Delta_m - 3}} \, Q^2\left(O_h O_m\right) + {\rm c.c.}~.
}}
Here we denote by $O$ the bottom component of the superfield $\CO$. In the
second line, we have traded the $\theta$-integral for the
action of the supercharges $Q^2$. Note
that this eliminates the extraneous numerical factors. At leading order, the interaction \spotlint\ gives the following contributions to the $B$-
and $A$-correlators:
\eqn\gmgmmassesi{\eqalign{
 & B(x) =  {   \lambda \vev{ Q^2 (O_h)}_h \over \Lambda^{\Delta_h + \Delta_m - 3}} \int d^4y \, \vev{ Q^2(O_m(y)) \,J(x) J(0)}_m~,\cr
 & A(x) =  {\lambda^2 \over  \Lambda^{2(\Delta_h + \Delta_m-3)}}
 \int d^4 y \, d^4 y' \,  \vev{Q^4\left(O_h^\dagger(y)  O_h(y')\right)}_h \cr
 & \hskip144pt \times \vev{Q^4\left(O_m^\dagger(y) O_m(y')\right) \, J(x) J(0)}_m~.
}}
Here we also use the subscripts $h$ and $m$ to highlight the factorization into SUSY-breaking correlators and messenger correlators.

We see that for superpotential interactions, the $B$-correlator is $\CO(\lambda)$ while the $A$-correlator is~$\CO(\lambda^2)$. Thus, the gaugino and sfermion masses will be of the same order in the interaction. This is the desired behavior expected of gauge mediation spectra. Note that global symmetries which are unbroken in either the SUSY-breaking sector or the messenger sector can make the correlators in \gmgmmassesi\ vanish. In particular, an unbroken $R$-symmetry in the messenger sector under which $R(\CO_m) \neq 2$ results in the vanishing of the leading-order $B$-correlator and hence of the gaugino mass.

\medskip
\medskip
\medskip

\centerline{\it General K\"ahler Potential Interactions}

\eqn\Kpotlint{\eqalign{
  \delta {\scr L} &= {16 \t \lambda \over \Lambda^{\t \Delta_h + \t \Delta_m -
2}} \int d^4 \theta \, \t \CO_h \t \CO_m + {\rm c.c.}  \cr
 &  =  {\t \lambda \over \Lambda^{\t \Delta_h + \t \Delta_m - 2}} \, Q^4
\left(\t O_h \t O_m\right) + {\rm c.c.} + (\hbox{total derivative})~.
 }}
 Now the leading-order $B$- and $A$-correlators are given by:
\eqn\gmgmmassesii{\eqalign{
 & B(x)  = { \t \lambda \vev{ Q^4(\t O_h)}_h \over \Lambda^{\t \Delta_h + \t \Delta_m - 2~}}   \int d^4 y \, \vev{Q^2\left(\tilde O_m(y)\right)\, J(x)J(0)}_m \cr
 & \hskip135pt + \left(\t O_{h,m} \rightarrow \t O_{h,m}^\dagger\right) + (\hbox{total~} x\hbox{-derivative})~,\cr
 &A(x) = {  \, \t \lambda \vev{ Q^4(\t O_h)}_h  \over \Lambda^{\t \Delta_h + \t \Delta_m -2}} \int d^4 y \, \vev{ Q^4\left(\t O_m(y)\right) \,J(x)J(0)}_m + {\rm c.c.}~.
}}
As we will discuss in section 4, this formula for $A(x)$ includes the well-known supertrace contribution to the sfermion mass-squared in weakly-coupled models of messengers interacting with $D$-term spurions \PoppitzXW.

Although the SUSY-breaking sector operator $\tilde\CO_h$ generically acquires
an $F$-term vev, this only leads to a total $x$-derivative in the
$B$-correlator~\gmgmmassesii. The reason is that the contribution of such an
$F$-term vev is proportional to the correlator
\eqn\zerocont{\vev{\t O_m(y) \, Q^4\left(J(x)J(0)\right)}_m~.}
Expanding $Q^4\left(J(x)J(0)\right)$ in components and using current
conservation shows that this correlator is a total $x$-derivative. Consequently,
it will not contribute to the gaugino mass~\ggmsoftmasses\ upon integrating the
$B$-correlator over~$x$. In section 3, we will give another, more conceptual
proof of this fact, which uses supersymmetric deformations and holomorphy. We
conclude that only the $D$-term vev of $\tilde\CO_h$ contributes to the leading-order gaugino mass. This fact has already been observed in examples, such as
\refs{\BuicanWS,\SeibergQJ}.

Just like for the superpotential contribution, an unbroken $R$-symmetry in the
messenger sector  can lead to a vanishing
$B$-correlator at leading order. However, in contrast to the
superpotential contribution, the $B$- and the $A$-correlators are now both
$\CO(\t\lambda)$. In general this will lead to a soft spectrum satisfying
$m_{\tilde f}\gg M_{\tilde g}~$, more fine-tuning in the SSM, and the
phenomenology of split SUSY \refs{\ArkaniHamedFB, \GiudiceTC}. The hierarchy $m_{\tilde f}\gg M_{\tilde g}$ can potentially be avoided, if the
$\CO(\t \lambda)$ contribution to the $A$-correlator vanishes (e.g.\ due to additional symmetries).

\medskip
\medskip
\medskip

\centerline{\it Half-Chiral K\"ahler Potential Interactions}

\medskip

These interactions are a subset of the previous case. They still have the form~\Kpotlint, except that now either $\tilde\CO_h$ or $\tilde\CO_m$ (but not both) is
chiral. In these cases we see from~\gmgmmassesii\ that the $\CO(\t \lambda)$
contribution to the $A$-correlator vanishes. The leading non-trivial contribution now arises at $\CO(\tilde\lambda^2)$.

When $\tilde\CO_h$ is chiral, we obtain:
\eqn\halfchsfm{\eqalign{A(x) = {{\t \lambda}^2 \over \Lambda^{2 (\t \Delta_h + \t \Delta_m - 2)}} & \int d^4 y \, d^4 y' \, \vev{Q^4\left(\t O_h^\dagger(y) \t O_h(y')\right)}_h \cr
 & \hskip49pt \times \vev{Q^4\left(Q^2(\t O_m^\dagger(y)) \b Q^2 (\t O_m(y'))\right) \, J(x)J(0)}_m~.}}
Note that by the discussion around \zerocont, a chiral $\t \CO_h$ can never give a gaugino mass at $\CO(\tilde\lambda)$, since there is no $D$-term vev, and even a nonzero $F$-term vev does not contribute to the $B$-correlator at zero momentum. In a sense, this is the most dramatic manifestation of the vanishing gaugino mass discussed above. Now the leading-order gaugino mass is $\CO(\t \lambda^2)$ and we again find a split-SUSY spectrum.\foot{It is straightforward to lower two powers of the interaction \Kpotlint\ and derive the $\CO(\t \lambda^2)$ contribution to the $B$-correlator; we do not display the result here.}

When $\t \CO_m$ is chiral, the leading-order $B$- and $A$-correlators are given by:
\eqn\halfchsfmii{\eqalign{&B(x) =  {   \t \lambda \vev{ Q^4 (\t O_h)}_h \over \Lambda^{\t \Delta_h + \t \Delta_m - 2}} \int d^4y \, \vev{ Q^2(\t O_m(y)) \,J(x) J(0)}_m~,\cr
&A(x) = {{\t \lambda}^2 \over \Lambda^{2 (\t \Delta_h + \t \Delta_m - 2)}}  \int d^4 y \, d^4 y' \, \vev{Q^4\left(Q^2(\t O_h^\dagger(y)) \b Q^2 (\t O_h(y'))\right)}_h \cr
 & \hskip166pt \times \vev{Q^4 \left( \t O_m^\dagger(y)\t O_m(y') \right) \, J(x)J(0)}_m~.}}
Note that in this case the $B$-correlator is $\CO(\t \lambda)$, while the $A$-correlator is $\CO(\t \lambda^2)$, so that the gaugino and sfermion masses are of the same order in the interaction.

\subsec{Simplifying Limits and their Phenomenology}

In formula~\gmgmmassesi, the leading-order $A$-correlator due to
superpotential interactions does not completely factorize into separate contributions from the SUSY-breaking sector and the messenger sector,
because of the momentum dependence of the SUSY-breaking correlator
\eqn\hcorr{
 \vev{Q^4\left(O_h^\dagger(y) O_h(y')\right)}_h~.}
In this subsection, we will discuss two scenarios in which this momentum
dependence becomes trivial and the $A$-correlator truly factorizes.\foot{An
analogous discussion applies to the hidden-sector correlators in \halfchsfm\ and
\halfchsfmii.} In these cases we will also be able to say more about the soft spectrum. The two simplifying limits are:

\medskip

 $\bullet$ {\it Small SUSY-Breaking: $F \ll M^2$.} Since the messenger correlator in \gmgmmassesi\ decays exponentially at long distance, we only need to consider the SUSY-breaking correlator~\hcorr\ at scales $|y - y'| \lesssim {1 \over M}\ll {1\over \sqrt{F}}$. The SUSY-breaking sector
at these scales is close to a fixed point.
Thus \hcorr\ can be approximated by the OPE of $O^\dagger_h$ and $O_h$ in this fixed-point CFT. The OPE translates into an
expansion in~$F/M^2$ and we can therefore focus on the leading operator
$O_\Delta$ in the OPE which acquires a SUSY-breaking $D$-term vev
$\vev{Q^4(O_\Delta)}_h \neq0$. Now formula~\gmgmmassesi\ for the $A$-correlator simplifies:
    \eqn\opefmla{\eqalign{&A(x) =
   {  \lambda^2 \vev{Q^4(O_\Delta)}_h \over \Lambda^{2(\Delta_h + \Delta_m-3)}} \int d^4 y \, d^4 y' \,  (y-y')^{\Delta-2\Delta_h} \vev{Q^4\left(O_m^\dagger(y) O_m(y')\right) J(x) J(0)}_m~.}}
Comparing \opefmla\ with the $B$-correlator in \gmgmmassesi, we see that the
ratio of the sfermion mass-squared to the gaugino mass-squared is given by:
\eqn\smallfratio{{m^2_{\t f} \over M^2_{\t g}} \sim \left({\sqrt{F} \over M}\right)^{\Delta-2 \Delta_h}~.}

There are three cases:

\medskip

\item{1.)} If $\Delta < 2 \Delta_h $, the sfermion mass is much larger then the
gaugino mass. The little hierarchy problem of the SSM is exacerbated, and the phenomenology is that of split SUSY \refs{\ArkaniHamedFB,
\GiudiceTC}.

\item{2.)} If $\Delta > 2 \Delta_h$, the sfermions are very light compared to
the gauginos. This is the interesting regime for the mechanism of hidden-sector
renormalization to operate. Indeed, our result~\smallfratio\ agrees
with~\refs{\CohenQC,\RoyNZ}.

\item{3.)} The boundary case $\Delta = 2 \Delta_h$ will typically produce
comparable sfermion and gaugino masses, just like ordinary gauge mediation.

\medskip

In a general CFT any of these options can in principle be realized. If the
SUSY-breaking sector is asymptotically free, we can say more: since the
product
$O_h^\dagger O_h$ is a good operator in the free CFT, we know that
$ \Delta\le 2\Delta_h$. Thus, we can either get sfermions which are much heavier
than the gauginos (1), or a spectrum of comparable sfermions and gauginos (3).
If $O_h$ is composite in the free UV regime, it is guaranteed that $\Delta<
2\Delta_h$
and we end up with split SUSY. If $O_h$ is a fundamental singlet we
get $\Delta=2\Delta_h=2$ and obtain a spectrum of comparable sfermions and
gauginos. This property of singlets is one reason they have played an important
role in model building (see e.g.\ \ShadmiMD).

\medskip

$\bullet$ {\it Spurion Limit.}  At sufficiently low energies, the dynamics of the SUSY-breaking sector becomes trivial, essentially containing only the vacuum energy and the Goldstino. Although we expect to find new states at the scale $\sqrt{F}$ (for instance, the scalar superpartner of the massless Goldstino), the dynamics of the theory could remain very weakly coupled and essentially classical through some much larger scale $\Lambda \gg \sqrt{F}$. This can happen in non-trivial examples, such as SUSY QCD with massive flavors below the strong-coupling scale \IntriligatorDD; another example will be discussed in subsection 4.3. Below the scale $\Lambda$, the only contribution to~\hcorr\ comes from picking the $F$-terms of the two operators; the correlator factorizes because the theory is classical:
\eqn\hcorrii{\vev{Q^4\left(O_h^\dagger(y) O_h(y')\right)}_h
= \left| \vev{Q^2(O_h)}_h\right|^2~.
    }
Again, formula~\gmgmmassesi\ for the $A$-correlator simplifies:
\eqn\spotsfm{\eqalign{ & A(x) =   {  \lambda^2 \left|\vev{Q^2(O_h)}_h\right|^2 \over  \Lambda^{2(\Delta_h + \Delta_m-3)}}
 \int d^4 y \, d^4 y' \,   \vev{\b Q^2 (O_m^\dagger(y)) Q^2 (O_m(y')) \, J(x) J(0)}_m~.}}
At scales larger than $\Lambda$, the dynamics of the SUSY-breaking sector becomes important and factorization no longer occurs. Since we cut off our effective theory at the scale $\Lambda$, this does not affect the result in \spotsfm.

This situation precisely corresponds to the well-known spurion regime in which
models of messengers are commonly studied.
Comparing \spotsfm\ to the $B$-correlator in \gmgmmassesi, we see that in this
regime the sfermion and gaugino masses are of the same order in SUSY-breaking,
so that we have
\eqn\relsizei{{m^2_{\t f} \over M^2_{\t g}} \sim 1~.}
This is expected from our experience with spurion models of messengers.

Note that if $F \ll M^2$ the spurion limit may overlap with the limit of small SUSY-breaking. In the spurion limit the dynamics of the SUSY-breaking sector is
trivial up to the scale $\Lambda$. For small SUSY-breaking this always
corresponds to case (3) discussed in the previous bullet point, and results in
comparable sfermion and gaugino masses, consistent with \relsizei. However, the
spurion limit may also apply in some cases where $F \gtrsim M^2$.

\newsec{Simplifying the Messenger Correlators}

The GMGM expressions for the $B$- and $A$-correlators derived in section 2 all
contain integrated products of SUSY-breaking correlators  $\vev{\ldots}_h$ and messenger correlators $\vev{\ldots}_m$. In this section, we
show how some of the supersymmetric messenger correlators can be
rewritten in terms of certain deformations of the original messenger Lagrangian.
This is particularly useful if the correlator coming from the SUSY-breaking
sector has trivial momentum dependence and factors out of the integral. In this
case, the $B$- and $A$-correlators can essentially be calculated  in a supersymmetric theory. We will use these results to elucidate certain aspects of the GMGM framework. They also lead
to a very compact treatment of weakly-coupled spurion models (see section 4 and
appendix B).

Consider the following supersymmetric deformation of the messenger Lagrangian:
\eqn\epdeformstart{\eqalign{
 \delta {\scr L}_{m}  & = {4 \epsilon \over \Lambda^{\Delta_m - 3}} \int d^2
\theta \, \CO_m +{16 \t \epsilon \over \Lambda^{\t \Delta_m - 2}} \int d^4
\theta \, \t \CO_m +{\rm c.c.}  \cr
&  =  {\epsilon \over \Lambda^{\Delta_m - 3}}\, Q^2\left(O_m\right) +{
\t\epsilon \over \Lambda^{\t \Delta_m - 2}} \, Q^4\left( \t O_m\right) +{\rm
c.c.} +(\hbox{total derivative})~.
}}
As before, both $\CO_m$ and $\t \CO_m$ are in general complex and appear in
\epdeformstart\ together with their complex conjugates. We also define the supersymmetric messenger correlator
\eqn\CSUSYdef{
 C_{\rm SUSY}(x;\epsilon,\tilde\epsilon) \equiv \vev{J(x)J(0)}_{m}~,
}
where the arguments $\epsilon$, $\t\epsilon$ indicate that this correlator is
to be evaluated in the messenger theory deformed by \epdeformstart.\foot{Note
that $C_{\rm SUSY}$ is the common supersymmetric limit of the functions $C_a$
defined in \MeadeWD\ (see appendix A).}

Derivatives of $C_{\rm SUSY}$ with respect to $\epsilon$, $\tilde\epsilon$
insert the operators appearing in \epdeformstart\ into the correlator on the
right-hand side of \CSUSYdef. As discussed above, this can be used to rewrite
various contributions to the leading-order $B$- and $A$-correlators, as follows:

\medskip

\item{1.)} Differentiating $C_{\rm SUSY}$ with respect to $\epsilon$ yields
\gmgmmassesi, the leading-order superpotential contribution to the
$B$-correlator:
\eqn\sdeformspgm{B(x) =  {  \lambda \vev{Q^2(O_h)}_h \over \Lambda^{\Delta_h}}
\, {\d \over \d \epsilon} C_{\rm SUSY}(x;\epsilon,\tilde\epsilon)
\biggr|_{\epsilon = \tilde\epsilon=0} ~.}
Formulas similar to \sdeformspgm\ were obtained in \refs{\DistlerBT,\IntriligatorFR} for examples of this type.

\item{2.)} Differentiating $C_{\rm SUSY}$ with respect to $\tilde\epsilon$
yields \gmgmmassesii, the leading-order K\"ahler potential contribution to the
$A$-correlator:
\eqn\kahlersfmdef{A(x) = { \t \lambda \vev{Q^4(\t O_h)}_h \over \Lambda^{\t
\Delta_h}} \, {\d \over \d \t \epsilon} C_{\rm SUSY}(x;\epsilon,\tilde\epsilon)
\biggr|_{\epsilon=\t \epsilon = 0} + {\rm c.c.}~.}

\item{3.)} A second derivative of $C_{\rm SUSY}$ with respect to $\epsilon$
yields \spotsfm, the leading-order superpotential contribution to the
$A$-correlator in the spurion limit:
\eqn\sfmdeform{A(x) = {\lambda^2 \left|\vev{Q^2(O_h)}_h\right|^2 \over
\Lambda^{2 \Delta_h}} \, {\d^2 \over \d \epsilon \d \epsilon^*} C_{\rm
SUSY}(x;\epsilon,\tilde\epsilon) \biggr|_{\epsilon =\tilde\epsilon= 0}~.}
This only holds in the spurion limit. Generically, the superpotential
contribution to the $A$-correlator cannot be written in terms of a
supersymmetric deformation, because of the non-trivial momentum dependence of
the SUSY-breaking two-point function $\vev{Q^4\left(O^\dagger_h(y) O_h(y')\right)}_h$ which appears in \gmgmmassesi.

\medskip
\medskip

Note that  the leading-order K\"ahler potential contribution \gmgmmassesii\ to
the $B$-correlator or the half-chiral contributions \halfchsfm\ and
\halfchsfmii\ to the $A$-correlator cannot be written as supersymmetric deformations of the messenger
Lagrangian.

We now give an alternate, more conceptual proof that the $F$-term vevs of
hidden-sector operators in the K\"ahler potential do not generate a leading-order
gaugino mass.\foot{An argument along these lines has been conjectured
in~\KomargodskiJF. The fact that $F$-term vevs in the K\"ahler potential cannot
generate leading-order gaugino masses may allow the discussion
in~\KomargodskiJF\ to be generalized further.}  Such a contribution would be proportional to the supersymmetric messenger correlator
\eqn\lala{
\vev{Q^4(\t O_m(y)) \, J(x)J(0)}_m \sim {\d \over \d \t \epsilon} C_{\rm SUSY}(x;\epsilon,\tilde\epsilon) \biggr|_{\epsilon=\t \epsilon = 0}~,
 }
evaluated at zero momentum. However, $C_{\rm SUSY}$ at zero momentum is (by definition) the wavefunction renormalization of the gauge multiplet at one-loop in the visible gauge coupling. In other words, the zero-momentum effective action for the gauge multiplet is given by
\eqn\lowenact{ {1 \over 4} \int d^2 \theta \, (1 + g^2\CC) W_\alpha^2 + \, {\rm c.c.}~,}
so that
\eqn\holrel{\t C_{\rm SUSY}(p = 0) = {\CC} + {\b \CC}~.}
Here $\t C_{\rm SUSY}(p)$ is the Fourier transform of $C_{\rm SUSY}(x)$. Thus, $\t C_{\rm SUSY}(p = 0)$ is the sum of a holomorphic function $\CC$ of the
microscopic hidden-sector couplings and an anti-holomorphic function $\b\CC$ of
these couplings. This splitting into holomorphic plus anti-holomorphic functions
is only true at zero momentum. Since $\t \epsilon$ is not a holomorphic
parameter, $\t C_{\rm SUSY}(p = 0)$ cannot depend on it, and hence the $\t
\epsilon$ derivative in \lala\ must vanish at zero momentum.

\newsec{Weakly-Coupled Examples}

In this section we explore a subset of the models we presented in
section~2: weakly-coupled spurion models of messengers.

\subsec{SUSY-Breaking in the Superpotential}

The most general renormalizable messenger theory with $N$ pairs $\Phi_i, \t \Phi_i$ of chiral messengers which couple to a spurion $X$ through the superpotential is given by:
\eqn\genmm{\eqalign{{\scr L}  = \int d^4 \theta \left(  \Phi^\dagger_i\Phi_i + \t \Phi^\dagger_i \t \Phi_i\right) + \int d^2 \theta \left(X \lambda_{ij} + m_{ij} \right)  \Phi_i \t \Phi_j \, + {\rm c.c.}~.}}
Here $\Phi_i, \t \Phi_i$ have $U(1)$ charges $+1$ and $-1$ respectively. The spurion $X$ acquires an $F$-term vev $X|_{\theta^2}=f~$ ($f$ real). This scenario was dubbed (Extra)Ordinary Gauge Mediation (EOGM) in \CheungES, where the phenomenology of such models was studied in detail (especially for the case where the theory possesses an $R$-symmetry).

In the language of section 2 we have (with an obvious generalization to multiple superpotential interactions):
\eqn\hiddnops{\CO_h = X~,\qquad \left(\CO_m\right)_{ij}=  \Phi_i \t \Phi_j~,}
with $\Delta_h = 1$ and $\Delta_m=2$. The couplings between $\CO_h$ and $\CO_m$ are $\sim \lambda_{ij}$. Since this model is renormalizable, the UV-cutoff $\Lambda$ does not appear in the interaction Lagrangian and the theory is fully calculable. For this model to be phenomenologically viable, we need to impose messenger parity (a symmetry that exchanges $\Phi$ and $\tilde \Phi$ under which the global $U(1)$ current is odd) and CP invariance. To satisfy these requirements, we assume that there exists a basis in which $m$ is real and diagonal with eigenvalues $m_i$, and that in this basis $\lambda$ is real and symmetric.

The GMGM framework tells us how to compute the leading-order contributions to the gaugino and sfermion masses: use the general formula \gmgmmassesi\ for superpotential interactions, with the operators \hiddnops. The result for the gaugino mass is:
\eqn\messgauginom{M_{\t g} = - {g^2 f \over 8 \pi^2} \sum_{i = 1}^N {\lambda_{ii} \over m_i}~,}
where $\lambda_{ii}$ are the diagonal entries of $\lambda$ in the basis where
$m$ is real and diagonal. For the sfermion mass-squared, we find:
\eqn\Afterms{m^2_{\t f} = {g^4 Y^2 f^2 \over 64 \pi^4} \sum_{i = 1}^N \left({\lambda_{ii}^2 \over m_i^2} + \sum_{j = 1 \atop j \neq i}^N {\lambda_{ij}^2 \over m_i^2 - m_j^2} \log {m_i^2 \over m_j^2}\right)~.}
These formulas are rederived in appendix B using the results of section 3.

Both \messgauginom\ and \Afterms\ were obtained in \CheungES\
using the wavefunction renormalization technique \GiudiceNI.
As we explain in appendix B, this technique is in general not applicable (even
for small SUSY breaking), but several
peculiarities of the EOGM Lagrangian \genmm\ render it valid in this particular
case.

An important property of the expression in \Afterms\ is that it is strictly
greater than zero. We have therefore shown that the leading-order sfermion
mass-squared in the most general renormalizable messenger model with $F$-term breaking is positive. In fact, we can prove a bound for the ratio of the sfermion mass-squared to the gaugino mass-squared. Since the second term on the right-hand side of~\Afterms\
above is
manifestly positive, we have \eqn\sfineqi{m^2_{\t f} \geq {g^4 Y^2
f^2 \over 64 \pi^4} \sum_{i=1}^N  {\lambda_{ii}^2 \over m_i^2}~.}
Using the fact that \eqn\CSineq{\sum_{i =1}^N {\lambda_{ii}^2
\over m_i^2} \geq {1 \over N} \left(\sum_{i = 1}^N {\lambda_{ii}
\over m_i}\right)^2~,}
we derive the inequality
\eqn\uoneineq{{m^2_{\t f} \over M^2_{\t g}}  \geq  {Y^2 \over N}~.}
It is clear that this inequality is saturated when all off-diagonal elements of
$\lambda_{ij}$ vanish and the ratio $\lambda_{ii} \over m_i$ is the same for
each messenger. For the simple case of a $U(1)$ visible gauge group considered
here, this is the usual definition of Minimal Gauge Mediation~(MGM) with $N$
messengers. Indeed, the ratio of the sfermion mass-squared to the gaugino mass-squared in such an MGM model is precisely $Y^2 \over N$. The inequality thus
states that renormalizable spurion models of weakly-coupled messengers with
$F$-term breaking can never give rise to scalars which are lighter than the ones
we get in MGM. This explains why such models are not sufficient to cover the
entire GGM parameter space \CarpenterWI.

Although we have only considered the case where the visible gauge group is
$U(1)$, it is clear that inequalities such as \uoneineq\ can be obtained for
messengers in arbitrary vectorlike representations of any gauge group. Our
discussion above is limited to leading order in SUSY-breaking.
It would be interesting to study the corrections to the inequality which arise at higher orders in
$f$ (in many examples these are known to be small). Other corrections come from renormalization group running in the
visible sector; we have not analyzed these effects in detail.

\subsec{SUSY-Breaking in the K\"ahler Potential}

We now consider the effect of coupling the messengers to a SUSY-breaking $D$-term spurion $V|_{\theta^4} = D~$ ($D$ real) through the K\"ahler potential. We start from the Lagrangian
\eqn\genmmii{{\scr L} = \int d^4 \theta \, \Phi^\dagger_i (\delta_{ij} + V \tilde\lambda_{ij}) \Phi_j + \t \Phi^\dagger_i (\delta_{ij} + V  \tilde\lambda_{ij}) \t \Phi_j + \int d^2 \theta \, m_{ij}  \Phi_i \t \Phi_j + {\rm c.c.}~.}
The matrix $\tilde\lambda$ must be Hermitian and messenger parity requires us to choose the same $\tilde\lambda$ for the $\Phi$ and the $\t \Phi$. In the language of section 2 (with an obvious generalization to multiple operators):
\eqn\gmgmkop{ \t \CO_h = V, \qquad  (\t\CO_m)_{ij} = \Phi_i^\dagger  \Phi_j + \t \Phi_i^\dagger \t \Phi_j~,
}
with couplings $\sim \t \lambda_{ij}$. Such terms do not by themselves contribute to the gaugino mass at any order in $D$ because of $R$-symmetry in the messenger sector. We will now analyze the sfermion mass-squared at leading order in $D$. After substituting \gmgmkop\ into \gmgmmassesii, we obtain:
\eqn\Adterms{m^2_{\t f} =  {g^4 Y^2 D \over 32 \pi^4} \sum_{i = 1}^N \tilde\lambda_{ii} \left(\log {\Lambda^2 \over m_i^2} - 2\right)~.}
Here $\tilde\lambda_{ii}$ are the diagonal elements of $\tilde\lambda$ in the basis in which $m$ is real and diagonal. The logarithmic divergence is a result of the fact that the SUSY-breaking $D$-term has introduced a non-vanishing supertrace $\sim \Tr \tilde\lambda$ into the messenger sector \PoppitzXW. To render the model calculable we assume that $\Tr \tilde\lambda = 0$  so that the supertrace vanishes.\foot{As emphasized in the introduction, we are assuming that there are no additional operators at the scale $\Lambda$ which can generate a sfermion mass.} This gives
\eqn\Adtermsii{m^2_{\t f} = - {\t g^4 Y^2 D \over 32 \pi^4} \sum_{i = 1}^N \tilde\lambda_{ii} \log m_i^2~.}
Note that as opposed to the $F$-term contribution \Afterms, this does not have definite sign. Formula \Adtermsii\ is rederived in appendix B using the results of section 3.

One could also have studied half-chiral terms in the K\"ahler potential, such as
\eqn\halfd{{1 \over \Lambda} \int d^4 \theta\,  X \Phi^\dagger \Phi + {\rm c.c.}~,}
where $X$ acquires an $F$-term vev $X|_{\thetasq} = f$, as in the previous
subsection. For simplicity, we now discuss a single pair of messengers ($N=1$)
of mass $m$.\foot{The fact that \halfd\ violates messenger parity is irrelevant
for this discussion.} We have shown in section~2, and
again in section~3, that such terms do not contribute to the
gaugino mass at leading order. In the language of this section,
this can be viewed as a consequence of the rescaling anomaly.\foot{The appearance of the rescaling anomaly in this example was already observed in \BuicanWS. Our discussion here serves to clarify its role in light of the general results of sections~2 and~3.} Redefining
\eqn\rescaling{\Phi \rightarrow \left(1-{X\over \Lambda}\right) \Phi~,}
the SUSY-breaking terms change according to
\eqn\rescalean{{1 \over \Lambda} \int d^4 \theta\,  X \Phi^\dagger \Phi + {\rm
c.c.} \rightarrow - {1\over \Lambda^2} \int d^4 \theta \, X^\dagger X
\Phi^\dagger \Phi - {m\over \Lambda} \int d^2 \theta \, X \Phi \t \Phi + {\rm
c.c.}~,}
where the second term on the right-hand side arises from the mass term $m\Phi\tilde\Phi$
in the superpotential. The terms in~\rescalean\
are a combination of $F$- and $D$-term spurions. Naively applying
\messgauginom,
the $F$-term leads to a gaugino mass
\eqn \wronggm{M_{\t g} = {g^2  \over 8 \pi^2} {f \over \Lambda}~.}
However, the rescaling \rescaling\ is anomalous and shifts
\eqn\resceff{W_\alpha^2 \rightarrow  \left( 1 + {g^2 \over 8 \pi^2} \log \left( 1 - {X\over \Lambda}\right) \right) W_\alpha^2 ~.}
This generates a contribution to the gaugino mass which exactly cancels
\wronggm. The role of the anomaly in this example is further elucidated in
appendix B. The anomaly
does not affect the sfermion mass, which can be obtained from \rescalean\ using
the formulas derived above for $F$- and $D$-term spurions.

\subsec{Semi-Direct Gauge Mediation}

Semi-Direct Gauge Mediation models \refs{\IbeWP\SeibergQJ-\ElvangGK} are concrete examples of weakly-coupled, completely
calculable gauge mediation.\foot{For earlier work along similar lines,
see \RandallZI.} They contain a SUSY-breaking sector and a messenger sector;
both sectors and the interactions which couple them can be treated in
perturbation theory. In this sense, they are described by the GMGM framework.
This framework, however, is most useful when the dynamics of the SUSY-breaking
sector and the messenger sector do not depend on the weak interaction between
them. We then expand to a given power in this interaction, but can in principle
treat the factorized SUSY-breaking and messenger correlators
exactly. This is not the case for semi-direct models, where the small coupling
between the SUSY-breaking sector and the messenger sector also plays a crucial
role in the dynamics of the SUSY-breaking sector itself (hence the name
``Semi-Direct Gauge Mediation''). In this case one must resort to ordinary
perturbation theory to also expand the SUSY-breaking correlator to the
desired order.

Note that once the SUSY-breaking vacuum has been fixed, these issues do not affect the first-order formulas derived in section~2 for general K\"ahler potential interactions. However, determining the higher-order contributions in general requires a full loop-calculation. These contributions are particularly important if the first-order contribution vanishes. We will see below that even in such cases, there is a parameter regime of Semi-Direct Gauge Mediation which can be analyzed using the results of the previous two subsections and appendix~B.

For concreteness, we specialize our discussion to the theory discussed in \SeibergQJ.
Here the SUSY-breaking sector is taken to be the $3$-$2$ model of
\AffleckXZ.\foot{The 3-2 model consists of an $SU(3) \times
SU(2)$ gauge theory with matter content resembling a single
generation of the standard model: $Q \in ( {\bf 3}, {\bf 2}); L
\in ({\bf 1},{\bf 2}); \b U, \b D \in (\b {\bf 3}, {\bf 1})$. The
numbers in parentheses label the $SU(3)$ and $SU(2)$
representations respectively of the matter fields. Assuming that
$\Lambda_3 \gg \Lambda_2$ (here $\Lambda_{2}, \Lambda_{3}$ are the
strong-coupling scales of the $SU(3),SU(2)$ gauge groups), the
$SU(3)$ dynamics dominates and the theory is described by an
effective superpotential \eqn\dynW{W_{\rm eff} = {\Lambda_3^7\over
\det (\b Q Q)} + h Q \b D L~,} where $\b Q = (\b U, \b D)$.} The parameters of the $3$-$2$ model are chosen to satisfy $h \ll g_2,
g_3 \ll 1$; we will denote $\alpha_2 =
{g_2^2 \over 4 \pi}$. There are $2N_f$ messengers $\ell_i \; (i = 1,\ldots, 2N_f)$, which transform as $SU(2)$-doublets. The messengers have a supersymmetric mass term given by
\eqn\Wtreesd{
W = m\ell^2~,
}
and couple to the SUSY-breaking sector through the $SU(2)$ gauge fields.

To compute the soft masses at first order in the interaction, we apply the general formulas  \gmgmmassesii\ from section~2. At this order, we see that the gaugino mass vanishes due to an $R$-symmetry in the messenger sector, while the sfermion mass vanishes because the $SU(2)$ generators are traceless. This requires us to compute beyond first order, which confronts us with all the difficulties discussed above.

There is, however, a parameter regime of Semi-Direct Gauge
Mediation in which the dynamics of the SUSY-breaking sector becomes
essentially trivial and the model reduces to a theory of weakly-coupled messengers interacting with spurions, like the theories
studied in the previous two subsections and in appendix B. The SUSY-breaking sector and the
messenger sector only interact through the $SU(2)$ gauge fields, which are completely Higgsed at the scale
\eqn\higgsscale{\Lambda \sim \Lambda_3/h^{1/7}~.}
Thus, the SUSY-breaking sector essentially becomes
trivial below this scale as far as the messengers are concerned. If we assume that $m \ll \Lambda$, the model reduces to an
effective theory of messengers interacting with spurions and
endowed with a natural UV-cutoff $\Lambda$. These
spurions are gauge invariant operators of the 3-2
model, and the messengers interact with them through an effective
K\"ahler potential (see section~3 of~\SeibergQJ). We can thus analyze this theory using the language of the previous two subsections, with spurions
giving rise to the following contributions to the messenger
spectrum~\SeibergQJ:

\medskip

\item{1.)} At zeroth order in $\alpha_2$, the SUSY-breaking sector gives the messengers diagonal mass-splittings~$M_{\rm mess}^2 \sim m^2 \pm m_d^2$ through a $D$-term spurion as in \genmmii:
\eqn\dterm{V|_{\theta^4} = D \sim m_d^2 \sim h^2 \Lambda^2~.}
At this order the
supertrace vanishes. The requirement that the messengers not be tachyonic restricts
$m^2 \gtrsim h^2 \Lambda^2$, so that the allowed parameter range for $m$ is
given by
\eqn\paramr{h \Lambda \lesssim m \ll \Lambda~.}

\item{2.) } At $\CO(\alpha_2)$, the 3-2 model generates a half-chiral K\"ahler potential interaction of the type displayed in \halfd\ with:
\eqn\halfd{X|_{\thetasq} = f \sim {\Lambda m_{od}^2 \over m}   \sim  { \alpha_2\over
4\pi}\, h \Lambda^2~.}
Here and below we will keep track of loop factors such as ${\alpha_2 \over 4\pi}$, but drop all other numerical $\CO(1)$ factors. As explained at
the end of the previous section, $f$ gives the messengers
off-diagonal masses $m_{od}^2 \sim {\alpha_2 \over 4\pi} h \Lambda m$. However, due to the rescaling
anomaly, no gaugino mass is generated at  $\CO(f)$.

\medskip

\item{3.)} A non-zero, negative supertrace for the messengers is also generated at $\CO(\alpha_2)$:
\eqn\str{{\rm Str} \, M_{\rm mess}^2 \sim  -N_f \alpha_2 h^2 \Lambda^2~.}

\medskip

We now discuss the leading non-trivial contributions of these three items  to the sfermion mass-squared. The $D$-term generates a negative sfermion
mass-squared at $\CO(D^4)$, while the $F$-term gives a positive contribution at $\CO(f^2)$. The supertrace \str\ gives rise to
a logarithmically divergent term $\sim - {\rm Str}
M^2_{\rm mess} \log \Lambda^2 / m^2$; this always dominates the
$\CO(f^2)$ contribution from the $F$-term, which we consequently
drop. We thus obtain for the sfermion mass-squared: \eqn\sfmstr{m^2_{\t f}
\sim Y^2\left({\alpha_g \over 4 \pi}\right)^2  N_f m^2
\left({h \Lambda\over m}\right)^2  \left(\alpha_2  \log {\Lambda^2\over
m^2}- C_{\t f} \left({h \Lambda \over m }\right)^6 \right)~,} where
$C_{\t f}$ is a positive $\CO(1)$ constant and $\alpha_g = {g^2
\over 4 \pi}$.

Because the supertrace leads to a log-divergent contribution which is cut
off at the scale $\Lambda$ at which we defined our spurion model, the
sfermion mass is UV-sensitive.  While this leading logarithmic
piece is universal and can be trusted, there are finite threshold
corrections at the scale $\Lambda$ which cannot be calculated in the
spurion limit we are considering. These corrections can be
estimated to be of the same order as the coefficient of the
leading logarithm, i.e. they are $\CO(\alpha_2)$. Since we limit ourselves to the
large-logarithm limit, these unknown threshold corrections can safely be
ignored. Note that in a certain regime, the threshold corrections may be
comparable to (or even dominate) the finite $D$-term contribution (the second term in \sfmstr). In that case, the latter can
also be ignored relative to the large logarithm. In other regimes, this negative $D$-term contribution can
dominate the threshold corrections; in that case, we must ensure that the log-enhanced term is sufficiently large to avoid a tachyonic sfermion mass.

It is important to note that the UV-sensitivity discussed in the previous
paragraph is an artifact of truncating the full theory to a spurion model. In
the example we are considering, the full theory is renormalizable and leads to
finite, calculable soft masses. Here the general discussion from the introduction applies: the more information we have about the
structure of the theory at the cutoff scale $\Lambda$, the more reliable
statements we can make about the soft masses.

Unlike the sfermion mass, the gaugino mass does not suffer from
a UV-ambiguity. As explained above, the $D$-terms by themselves never generate a gaugino mass, and the $\CO(f)$ contribution vanishes due to the rescaling anomaly. Using the formulas from appendix B, we see that
the leading non-trivial contributions to the gaugino mass arise at \hskip25pt $\CO(f D^2)+\CO(f^3)$: \eqn\semigaugino{M_{\t g} \sim \left({\alpha_g\over
4\pi}\right)\left({\alpha_2 \over4\pi}\right) \,  \, N_f m
\left({h\Lambda\over m}\right)^3 \, \left(\left({h\Lambda \over
m}\right)^2 + C_{\t g}
\left({\alpha_2\over4\pi}\right)^2\right)~,} where $C_{\t g}$ is a
positive $\CO(1)$ constant. Parametrically either the first or the
second term dominates, depending on the relative size of ${
\alpha_2\over 4\pi}$ and ${h \Lambda \over m}$.

We now compare the sfermion mass-squared to the gaugino mass-squared:
\eqn\sfmgrat{{m_{\t f}^2 \over M_{\t g}^2} \sim {4\pi
Y^2 N_f \log{\Lambda^2\over m^2} \over \left({\alpha_2\over
4\pi}\right)\left({h\Lambda\over m}\right)^4\left(\left({h\Lambda\over
m}\right)^2+C_{\t g} \left({\alpha_2\over 4\pi}\right)^2 \right)^2} ~.} For the parameter range \paramr\ and $\alpha_2 \ll 1$, the sfermion mass is always much greater than
the gaugino mass. (This can be avoided by fine-tuning the hidden-sector parameters.) Thus, the
phenomenology is that of split SUSY, and the fine-tuning problem
of the SSM is exacerbated. This does not rule out the possibility
of a more desirable phenomenology for the case $m \gtrsim \Lambda$,
where the spurion treatment breaks down and a genuine loop-calculation is required.

\bigskip
\bigskip
\noindent {\bf Acknowledgments:}

We would like to thank A.~Giveon, K.~Intriligator, D.~Marques, Y.~Nakai, and Y.~Ookouchi  for
useful discussions. The work of TD was supported in part by NSF
grant PHY-0756966 and a Centennial Fellowship from Princeton
University. The work of ZK was supported in part by NSF grant
PHY-0503584. The work of NS was supported in part by DOE grant
DE-FG02-90ER40542. The work of DS was supported in part by DOE
grant DE-FG02-90ER40542 and the William D. Loughlin membership at
the Institute for Advanced Study. Any opinions, findings, and
conclusions or recommendations expressed in this material are
those of the author(s) and do not necessarily reflect the views of
the National Science Foundation.

\appendix{A}{Remarks on General Gauge Mediation}

In this appendix we briefly review the formalism of General Gauge Mediation (GGM) and collect some results on the GGM correlation functions which have not yet been discussed in the literature.

\subsec{Review of GGM}

The GGM formalism applies to theories which decouple into a SUSY-breaking hidden sector and a supersymmetric visible sector when the visible-sector gauge couplings vanish. For simplicity, we will take the visible sector to consist of a $U(1)$ gauge theory with coupling~$g$ and a single flavor~$f$ of charge~$Y$. Of central importance are the correlation functions of the hidden-sector global $U(1)$ current multiplet. A global symmetry current~$j_\mu$ is embedded in a real superfield $\CJ$ satisfying $D^2\CJ=0$. In components:
\eqn\currsf{\CJ = J + i \theta j - i \thetabar \b j - \theta \sigma^\mu \thetabar j_\mu + \half \thetasq \, \thetabar \sigmabar^\mu \d_\mu j - \half \thetabarsq \, \theta \sigma^\mu \d_\mu \b j - {1 \over 4} \thetasq \thetabarsq\,  \d^2J~,}
with $\d_\mu j^\mu = 0$.

We define the functions $C_a(x) \; \; (a = 0, {1/2}, 1)$ and $B(x)$ through
\eqn\Cfunsdef{\eqalign{ & \vev{J(x)J(0)}  = C_0(x)  \cr
 &   \vev{j_\alpha(x) \b{j}_\alphadot(0)} = -i \sigma^\mu_{\alpha\alphadot} \d_\mu C_{1/2}(x)  \cr
&  \vev{j_\mu(x) j_\nu(0)} = (\eta_{\mu\nu}\d^2 - \d_\mu \d_\nu) \, C_1(x) \cr
 &  \vev{j_\alpha(x) j_\beta(0)} =  {1 \over 4}\,  \ep_{\alpha\beta}  B(x) ~.}}
Note that our normalization of $B$ differs from the conventions in \refs{\MeadeWD,\BuicanWS} by a factor of~$4$. The Fourier transforms of the functions $C_a, B$ will be denoted by $\tilde C_a,\tilde B$.\foot{This differs from the notation in \refs{\MeadeWD,\BuicanWS} by some powers of $x$. We adopt this convention so that we can consistently denote by $\t f(p)= \int d^4 x \, f(x) e^{-ipx}$ the Fourier transform of a function~$f(x)$, while matching the conventions of \refs{\MeadeWD,\BuicanWS} in momentum space.} When SUSY is unbroken, all the $C_a(x)$ are equal, and we denote their common limit by $C_{\rm SUSY}(x)$.
The SUSY-breaking gaugino mass and sfermion mass-squared are then given by:
\eqn\gauginomass{M_{\t g} = {g^2 \over 4} \t B(p = 0)~,}
\eqn\sfmass{m^2_{\t f} = -g^4 Y^2 \int {d^4 p \over (2\pi)^4} \, {1 \over p^2} \, \left( \t C_0(p) - 4 \t C_{1/2}(p) + 3 \t C_1(p)\right)~.}
Finally, recall from~\BuicanWS\ that we can write
\eqn\Qfmlai{B(x) = \vev{Q^2 \left(J(x) J(0)\right)}~,}
\eqn\Qfmlaii{A(x) \equiv -8 \d^2 \left(C_0(x) - 4 C_{1/2}(x) + 3 C_1(x)\right) = \vev{Q^4\left( J(x) J(0)\right)}~,}
where $Q^2(\cdots) = \{Q^\alpha, \left[Q_\alpha, \cdots \right] \}$ and $Q^4(\cdots) = \{Q^\alpha, [Q_\alpha, \{\b Q_\alphadot, [\b Q^\alphadot, \cdots]\}]\}$. The order of the supercharges in~\Qfmlai\ and~\Qfmlaii\ is inconsequential because of translational invariance. Using \Qfmlaii\ and integrating by parts, formula \sfmass\ can be put into the form \ggmsoftmasses.

\subsec{Goldstinos in GGM}

Supersymmetric theories without FI-terms or non-trivial target-space geometry in the UV have a Ferrara-Zumino multiplet~\refs{\FerraraPZ\KomargodskiRB-\KuzenkoAM} containing the supercurrent and the energy-momentum tensor. It is organized in terms of a real vector superfield $\CJ_{\alpha\alphadot}$ satisfying
\eqn\FZdef{\b  D^\alphadot \CJ_{\alpha\alphadot}=D_\alpha  {\bf X}~,\qquad \b D_\alphadot {\bf X}=0~.}
We see that there must be a chiral superfield ${\bf X}$ which is well-defined in all supersymmetric theories we discuss. In~\KomargodskiRZ, it was shown that at low energies the operator ${\bf X}$ flows to an operator $X_{NL}$ as follows:
\eqn\flowUVIR{\eqalign{&{\bf X}\rightarrow {8F\over 3}X_{NL}~,\cr&
X_{NL}={G^2\over 2F_X}+\sqrt2\theta G+\theta^2 F_X~.}}
Here $G$ is the massless Goldstino fermion; $X_{NL}$ satisfies $X_{NL}^2=0$. At very low energies, $F_X$ can be replaced by its expectation value $F$, where $F^2$ is the vacuum energy-density.

Consider a general superfield $\CO$, which is well-defined in the UV. Suppose that the expectation values of its $\thetasq,\thetabarsq,\theta^4$ components are $F_\CO, G_{\CO},D_{\CO}$, respectively. At very low energies this superfield must flow to
\eqn\comb{\CO\rightarrow {F_\CO\over F}X_{NL}+{G_\CO\over F}X^\dagger_{NL}+{D_\CO\over F^2} X_{NL}^\dagger X_{NL}+\cdots~,}
where the dots stand for corrections with more fermions or more derivatives. If there are other massless particles, they could mix into~\comb\ as well, but this will not change our final answer.

We can use the decomposition~\comb\ to extract the $F,G,D$-components of the superfield~$\CO$ by projecting onto combinations of $X_{NL}$, $X_{NL}^\dagger$. Equivalently, we can consider correlation functions of $\CO$ and combinations of $\bf X$, $\bf X^\dagger$ at very long distances as follows:
\eqn\corrfun{\eqalign{&F_\CO={3\over 2}\pi^4 F^2\lim_{y\rightarrow\infty}y^6\langle  X^\dagger(y) \CO(0) \rangle~,\cr & G_\CO={3\over 2}\pi^4 F^2\lim_{y\rightarrow\infty}y^6\langle  X(y) \CO(0) \rangle~,\cr
& D_\CO={9\over 4}\pi^8 F^4\lim_{y\rightarrow\infty}y^{12}\langle  X^\dagger X(y) \CO(0) \rangle~.
}}
Here $X$ denotes the bottom component of the superfield $\bf X$. The last equation in~\corrfun\ contains no self-contractions between the $X$ operators. This is evident from the expansion~\comb.

These observations allow us to rewrite the $B$- and $A$-correlators \Qfmlai\ and \Qfmlaii\ as
\eqn\softmasses{\eqalign{
& B(x) = \vev{Q^2 (J(x) J(0))} = 6\pi^4 F^2\lim_{y\rightarrow\infty}y^6\langle  X^\dagger(y) J(x)J(0) \rangle
~,\cr
& A(x) = \vev{Q^4 (J(x) J(0))} = 36\pi^8 F^4\lim_{y\to\infty}y^{12}\langle X^\dagger X(y)J(x)J(0)\rangle~.
}}
This converts the GGM formulas to correlation functions of bottom components of well-defined superfields, without the need to perform SUSY variations. Physically the first formula in~\softmasses\ describes two Goldstinos $G^2$ propagating from infinity at essentially zero momentum and contracting with $J(x)J(0)$. Similarly the second formula describes four zero-momentum Goldstinos $G^2 \b G^2$ coming from infinity.

\appendix{B}{More on Weakly-Coupled Messengers}

In this appendix, we give a more detailed treatment of the weakly-coupled
spurion models discussed in section 4.

\subsec{SUSY-Breaking in the Superpotential}

The method of section 3 tells us how to compute the visible soft masses for the
messenger Lagrangian
\eqn\genmm{\eqalign{{\scr L}  = \int d^4 \theta \left(  \Phi^\dagger_i\Phi_i + \t \Phi^\dagger_i \t \Phi_i\right) + \int d^2 \theta \left(X \lambda_{ij} + m_{ij} \right)  \Phi_i \t \Phi_j \, + {\rm c.c.}~,}}
with $X|_{\thetasq} = f$ (see subsection 4.1), by studying the {\it supersymmetric} Lagrangian
\eqn\susylag{{\scr L_{\rm SUSY}} =\int d^4 \theta \left(\Phi^\dagger_i \Phi_i + \t \Phi_i^\dagger \t \Phi_i\right) + \int d^2 \theta \,  \CM_{ij} \Phi_i \t \Phi_j + \, {\rm c.c.}~. }
Here $\CM$ is a general complex supersymmetric mass matrix that should be distinguished from $m_{ij}$.

Because of $SU(N)\times SU(N)$ invariance, all physical observables depend only on the eigenvalues $\mu_i^2$ of $\CM$. Thus, the function $\t C_{\rm SUSY}(p)$ takes the simple form
\eqn\csusyiii{\t C_{\rm SUSY}(p) = {1 \over 8 \pi^2} \sum_{i = 1}^N \left(\log {\Lambda^2 \over \mu_i^2} + g \left({p^2 \over \mu_i^2}\right)\right)~,}
where $\Lambda$ is a UV-cutoff. The numerical coefficient and the functional form of $g$ are fixed by a one-loop calculation. We will only need the asymptotic behavior
\eqn\asymtotics{\eqalign{g(p^2 \rightarrow 0) = -1 + \CO(p^2)~, \cr
g(p^2 \rightarrow \infty) = - \log { p^2 \over \mu_i^2} + 1 + \CO(1 / p^2)~.}}
Substituting this into \csusyiii, we see that
\eqn\zeromomcsusy{\t C_{\rm SUSY}(p = 0) = {N \over 8 \pi^2} \left(\log {\Lambda^2} -1\right) - {1 \over 8 \pi^2} \left( \Tr \log  \CM + \Tr \log \CM^\dagger  \right)~.}
Thus, at zero momentum the answer breaks up into a holomorphic and an anti-holomorphic part, as was explained in section 3.

It is now straightforward to calculate the soft masses. In the language
of sections 3 and 4, we consider the deformation $\CM = m + \epsilon \lambda$ in
\zeromomcsusy\ and differentiate as in \sdeformspgm\ to obtain the gaugino mass:
\eqn\messgauginom{M_{\t g} = - {g^2 f \over 8 \pi^2} {\d \over \d \epsilon} \Tr
\log (m + \epsilon \lambda) =- {g^2 f \over 8 \pi^2} \Tr\,m^{-1}\lambda = - {g^2
f \over 8 \pi^2} \sum_{i = 1}^N {\lambda_{ii} \over m_i}~,}
where $m_i$ are the eigenvalues of $m$ and $\lambda_{ii}$ are the diagonal entries of $\lambda$ in the basis where $m$ is real and diagonal.

To calculate the sfermion mass-squared, we need to calculate a second derivative
of $\t C_{\rm SUSY}(p)$ with respect to $\epsilon$ and $\epsilon^*$, as in
\sfmdeform.
The result then needs to be integrated as in \ggmsoftmasses\ (or equivalently \sfmass). For this class of
theories, there is a trick to perform the integral.\foot{In some special cases
this has been discussed in \refs{\DistlerBT,\IntriligatorFR}. } Because of the
simple form of $\t C_{\rm SUSY} (p)$ in \csusyiii, the $\epsilon$-derivatives
can be converted to momentum derivatives using the chain rule:
\eqn\epderiv{{\d^2 \over \d \epsilon \d \epsilon^*} \t C_{\rm SUSY}(p) = -
{p^2\over 8 \pi^2} \sum_{i = 1}^N \left({\d^2 \log \mu_i^2 \over \d \epsilon \d
\epsilon^*} {\d g \over \d p^2} - \left|{\d \log \mu_i^2 \over \d
\epsilon}\right|^2 {\d \over \d p^2} \left( p^2 {\d g \over \d
p^2}\right)\right)~.}
It is now straightforward to integrate, picking up the boundary values according
to \asymtotics. We obtain for the sfermion mass-squared:
\eqn\Afterms{\eqalign{m^2_{\t f} & = {g^4 Y^2 f^2 \over 64 \pi^4} \sum_{i=1}^N
\left({\d^2 \log \mu_i^2 \over \d \epsilon \d \epsilon^*} \, \log \mu_i^2 +
\left| {\d \log \mu_i^2 \over \d \epsilon} \right|^2\right)  \cr
& = {g^4 Y^2 f^2 \over 128 \pi^4} {\d^2  \over \d \epsilon \d \epsilon^*} \Tr
\log^2 \CM^\dagger \CM\cr
 & = {g^4 Y^2 f^2 \over 64 \pi^4} \sum_{i = 1}^N \left({\lambda_{ii}^2 \over m_i^2} + \sum_{j = 1 \atop j \neq i}^N {\lambda_{ij}^2 \over m_i^2 - m_j^2} \log {m_i^2 \over m_j^2}\right)~.}}

In \messgauginom\ and \Afterms\ we find perfect agreement with the answers quoted in subsection 4.1. These results were first obtained in~\CheungES\ using the wavefunction renormalization technique~\GiudiceNI. In the next subsection, we will reexamine this technique in more detail, comment on its limitations (even for small SUSY-breaking), and explain why it happens to give correct answers for the soft masses in weakly-coupled spurion models.

\subsec{Comments on Wavefunction Renormalization}

We begin by reviewing the wavefunction renormalization technique \GiudiceNI\ for the
sfermion mass in Minimal Gauge Mediation (MGM). Consider a single pair of
messengers $\Phi, \t \Phi$ with superpotential
\eqn\weqn{W = X \t \Phi \Phi~.}
For now $X$ is a background chiral superfield and SUSY is unbroken. To obtain
the sfermion mass, we need to calculate the $X$-dependent supersymmetric
effective action for the sfermion fields $Q, \t Q$. There are
two types of operators which can appear in this effective action: operators which contain the UV-cutoff $\Lambda$ and
operators which are $\Lambda$-independent. The only place where $\Lambda$ can
appear is inside perturbation theory logarithms; the first non-trivial such operator appears at two-loop order and contributes to the anomalous dimension of $Q, \t Q$. In MGM it
is given by:
\eqn\mgmgkahler{\delta K \sim \log^2 {X^\dagger X \over \Lambda^2}\,
\left(Q^\dagger Q + \t Q ^\dagger \t Q\right)~.}
Now consider $\Lambda$-independent terms. We organize the effective action as an
expansion in the number of supercovariant derivatives acting on $X, X^\dagger$.
Since $X$ is the only mass scale, there is
no operator without covariant derivatives. There are, however, many operators
with covariant derivatives, such as
\eqn\covderivops{\int d^4 \theta \,  {\b D^2 X^\dagger D^2 X \over
(X^\dagger X)^2}\, \left(Q^\dagger Q + \t Q ^\dagger \t Q\right)~.}

To introduce SUSY-breaking, we give the $F$-term of $X$ a vev $X|_{\thetasq} =
f$. The contribution of \mgmgkahler\ and \covderivops\ to the sfermion
mass-squared is (up to coefficients) given by:
\eqn\sqmass{m_{\t f}^2 \sim  {f^2 \over |X|^2} + {f^4 \over |X|^6}~.}
In this equation $X$ denotes the bottom-component vev of the
background superfield $X$. The wavefunction renormalization technique correctly
captures the $\CO(f^2)$ term in~\sqmass\ because the only operator~\mgmgkahler\
which can be written at this order is the~cutoff-dependent anomalous dimension.
The higher orders are not captured and are more difficult to calculate.

This discussion suggests that the wavefunction renormalization technique may not
even capture the  leading SUSY-breaking contribution in theories with more than
one mass scale. Consider, for example, an O'Raifeartaigh-like model with
superpotential
\eqn\cwmode{W = \CM(X)_{ij}  \Phi_i \t \Phi_j + Y \rho_{ij} \Phi_i \t \Phi_j,}
where $\CM(X)$ is a general matrix function of the background chiral
superfield $X$ and the K\"ahler potential is canonical. Suppose we want to
calculate the one-loop effective potential for $Y$ when $X$ acquires a
SUSY-breaking vev $X|_{\thetasq} = f$. It is easy to check that the wavefunction
renormalization technique fails to give the correct answer even at leading order
in $f$. The reason is that using the matrices $\CM(X)$ and $\rho$ we can
construct many cutoff-independent operators without covariant derivatives. To see this, consider the effective K\"ahler potential generated for $Y$ when the
chiral fields $\Phi, \t \Phi$ are integrated out at one-loop; it is given by
\IntriligatorDD:
\eqn\operator{K_{\rm eff} \sim \Tr \left(\left(\CM + \rho Y\right) ^\dagger
\left(\CM + \rho Y\right) \log {\left(\CM + \rho Y\right)^\dagger \left(\CM +
\rho Y\right) \over \Lambda^2}\right)~.}
Expanding the logarithm, we obtain many $\Lambda$-independent operators
without covariant derivatives (for instance, a tadpole for $Y$).  Therefore, the wavefunction renormalization technique does not correctly capture
the leading-order SUSY-breaking effect. We generally expect this to be the case,
unless the theory has essentially only one scale set by a single superfield.

The example of MGM discussed above exactly falls into this class of trivial
theories. More generally, consider a free theory of $N$
messengers of the type discussed in subsections~4.1 and~B.1, with superpotential
\eqn\weqnii{W = (X \lambda_{ij}  + m_{ij} ) \Phi_i \t \Phi_j = \CM_{ij}(X)
\Phi_i \t \Phi_j~.}
The effective action for this theory can only depend on the eigenvalues $\mu_i^2$ of the matrix $\CM(X)$. Since the
messengers are decoupled and the gauge interactions are flavor blind, ratios of
different eigenvalues cannot appear. Thus, the same argument as for MGM shows
that there are no cutoff-independent operators at leading order in $f$, and at
that order the result is correctly captured by the wavefunction renormalization
technique. Note that the assumption that the messengers are free is crucial for
this argument.

\subsec{SUSY-Breaking in the K\"ahler Potential}

Proceeding as in subsection B.1, we now compute the sfermion mass-squared
for the messenger Lagrangian
\eqn\genmmii{{\scr L} = \int d^4 \theta \, \Phi^\dagger_i (\delta_{ij} + V \tilde\lambda_{ij}) \Phi_j + \t \Phi^\dagger_i (\delta_{ij} + V \tilde\lambda_{ij}) \t \Phi_j + \int d^2 \theta \, m_{ij}  \Phi_i \t \Phi_j + {\rm c.c.}~,}
with $V|_{\theta^4} = D$ and $\t \lambda$ Hermitian, by analyzing the SUSY theory given by
\eqn\susylagii{{\scr L_{\rm SUSY}} = \int d^4 \theta \, \Omega_{ij}(\Phi^\dagger_i \Phi_j + \t \Phi^\dagger_i \t \Phi_j) + \int d^2 \theta \, m_{ij} \Phi_i \t \Phi_j + {\rm c.c.}~.}
Here $\Omega$ is Hermitian positive-definite. As discussed in subsection 4.2, the Lagrangian~\genmmii\ by itself never gives rise to gaugino masses because of $R$-symmetry.

We want to compute $\t C_{\rm SUSY}(p)$, as a function of the deformation
$\Omega = \1 + \t \epsilon\,  \tilde\lambda$, and then take a derivative with
respect to $\t \epsilon$ as in \kahlersfmdef.
To calculate $\t C_{\rm SUSY}(p)$, we first diagonalize~$\Omega$ through a unitary transformation:
\eqn\diageqn{U^\dagger \Omega U = {\rm diag}( | \omega_i|^2)~,}
where the $\omega_i$ are unique up to a phase. This $SU(N)$ transformation is not anomalous, and we can rewrite the Lagrangian as
\eqn\susylagii{{\scr L_{\rm SUSY}} = \sum_{i = 1}^N \int d^4 \theta \,|\omega_i|^2 \left( \Phi^\dagger_i\Phi_i +  \t \Phi_i^\dagger \t \Phi_i\right) + \int d^2 \theta \, m'_{ij}  \Phi_i \t \Phi_j + \, {\rm c.c.}~,}
where $m' = U^{T} m U$. We now rescale $\Phi_i \rightarrow \omega_i^{-1} \Phi_i$, $\t \Phi_i \rightarrow \omega_i^{-1} \t \Phi_i$ and obtain
\eqn\susylagiii{{\scr L_{\rm SUSY}}  = \sum_{i = 1}^N \int d^4 \theta \,\left(\Phi^\dagger_i\Phi_i +  \t \Phi_i^\dagger \t \Phi_i\right) + \int d^2 \theta \, m_{ij}'' \t \Phi_i \Phi_j + \, {\rm c.c.}~, }
where $m'' = {\rm diag} (\omega_i^{-1})\,  m' \, {\rm diag} ( \omega_i^{-1})$. However, this rescaling is anomalous and we pick up a correction of the form
\eqn\anomaly{W_\alpha^2 \rightarrow \left(1 - {g^2\over 4 \pi^2} \sum_{i = 1}^N \log \omega_i \right) W_\alpha^2~.}

Note that the eigenvalues $\mu_i^2$ of $m''^\dagger m''$ are the physical masses of the messengers in this supersymmetric theory; they depend on the matrix $\Omega$ and its eigenvalues $|\omega_i|^2$ in a complicated way. Since we have analyzed models of the type \susylagiii\ in subsection B.1, we only need to add the contribution of the anomaly to obtain
\eqn\csusyiv{\t C_{\rm SUSY}(p) = {1 \over 8 \pi^2} \sum_{i = 1}^N \left(\log {\Lambda^2 \over \mu_i^2} + g \left({p^2 \over \mu_i^2}\right)- 2 \log | \omega_i|^2 \right)~.}
Here $g$ is the same function as in \csusyiii. It is easy to check that at zero momentum \csusyiv\ breaks up into  the sum of a holomorphic and an anti-holomorphic part, in accordance with the discussion in section 3. In particular, $\t C_{\rm SUSY}(p = 0)$  is $\Omega$-independent.

We now set $\Omega = \1 + \t \epsilon \,\t\lambda$ and differentiate with
respect to $\t \epsilon$. Since $\t C_{\rm SUSY}$ is independent of $\Omega$ at
zero momentum, only $g$ can contribute to this derivative. Converting $\t
\epsilon$-derivatives to momentum derivatives as before, we obtain:
\eqn\eptildederiv{{\d \over \d \t \epsilon} \t C_{\rm SUSY}(p) = - {p^2 \over 8 \pi^2} \sum_{i = 1}^N {\d \log \mu_i^2 \over \d \t \epsilon} {\d g \over \d p^2}~.}
Performing the integral is again trivial, and we get:
\eqn\Adterms{m^2_{\t f} =  {g^4 Y^2 D \over 32 \pi^4} \sum_{i = 1}^N \tilde\lambda_{ii} \left(\log {\Lambda^2 \over m_i^2} - 2\right)~.}
Here $\tilde\lambda_{ii}$ are the diagonal elements of $\tilde\lambda$ in the basis in which $m$ is real and diagonal. As discussed in subsection 4.2, we assume that $\Tr \tilde\lambda = 0$  to ensure a vanishing supertrace and render the model calculable. This finally gives
\eqn\Adtermsii{m^2_{\t f} = - {\t g^4 Y^2 D \over 32 \pi^4} \sum_{i = 1}^N \tilde\lambda_{ii} \log m_i^2~,}
in accordance with our result from subsection 4.2.

\subsec{Comments on Vanishing Gaugino Masses}

In this subsection we elaborate on the discussion at the end of subsection 4.2 and elucidate why chiral spurions in the K\"ahler potential cannot generate leading-order gaugino masses.

Consider a single free chiral superfield $\Phi$. The theory
has a global $U(1)$ symmetry with a conserved current
$\CJ=\Phi^\dagger\Phi$.  The anomaly is the statement that while we can satisfy
\eqn\threenzds{D_i^2 \vev{ \CJ(x_1,\theta_1, \bar \theta_1)
\CJ(x_2,\theta_2, \bar \theta_2) \CJ(x_3,\theta_3, \bar \theta_3)
}=0\qquad (i = 1, 2, 3)}
at separated points, there is no set of contact terms which makes \threenzds\ true at coincident points.
One way to cancel this anomaly is to introduce an additional
chiral superfield $\tilde \Phi$  with its own current $\tilde \CJ=
\tilde \Phi^\dagger \tilde \Phi$.  This makes it possible to choose contact terms such that the current $\hat \CJ=\CJ-\tilde \CJ$ satisfies $D^2 \hat \CJ=0$ in all correlation
functions -- even at coincident points.

Perhaps the most dramatic consequence of these contact terms is
that even though the fields $\Phi$ and $\tilde \Phi$ are
decoupled, we must have \eqn\conse{\vev{ \CJ \CJ \CJ}
\not= \vev{ \CJ \hat \CJ \hat \CJ}~.} The two correlators
in \conse\ differ by contact terms at coincident points.  No Feynman
diagram with intermediate $\Phi$ or $\tilde \Phi$ fields leads to such
contact terms.  However, if we add heavy regulator fields, then the contact terms are generated by diagrams with intermediate regulator particles. This phenomenon is very common in the presence of
anomalies.

We now apply these statements to study the gaugino mass for a single messenger pair~$\Phi, \t \Phi$ interacting with a chiral spurion $X$ through the K\"ahler potential.  For now, we assume that the messengers are massless, so that in the notation of this subsection the Lagrangian is given by
\eqn\lagt{{\scr L} = \int d^4 \theta \left(\tilde \CJ + \left(1 + {X \over \Lambda} +
{X^\dagger \over \Lambda} \right) \CJ\right)~.}
To leading order in $X|_{\thetasq} = f$, the gaugino
mass is determined by integrating the correlator
\eqn\gaum{b(x_1, x_2, x_3) \sim  \vev{ Q^4 (J(x_1)) \hat J (x_2) \hat
J (x_3)}~.} Note that $D^2 \CJ=0$ implies that  $Q^4 J =0$ and therefore $b$ vanishes at
separated points. However, it does not vanish identically: there are finite contact terms at coincident points.  Consequently, the massless theory
based on~\lagt\  leads to a gaugino mass. This mechanism is identical to the one discussed in~\DineME.

If we add a supersymmetric mass term \eqn\suppoa{W=m \Phi \tilde \Phi~,} with
arbitrary nonzero $m$, then there is an additional contribution to the
correlation function $b$. The reason is that the current $\CJ$ is no longer
conserved, and thus $b$ no longer vanishes at separated points. However, when integrated, the contribution from separated points exactly cancels the contact-term contribution from coincident points. As was already discussed at the end of subsection 4.2, this can be directly seen by rescaling $\Phi \to
\left(1- {X \over \Lambda}\right)\Phi$.  Classically, this leads (at leading order in $X$) to canonical kinetic terms and an
$X$-dependent superpotential, which would seem to generate a gaugino
mass. However, quantum mechanically the rescaling is
anomalous, and the effect of the anomaly precisely cancels the superpotential contribution to the gaugino mass. This is consistent with our
arguments from sections 2 and 3 based on the explicit evaluation of
correlation functions and holomorphy.

In summary, we see that for $m=0$ the anomaly leads to a nonzero gaugino mass even though there is no obvious diagram in the low-energy theory which generates such a mass.  For nonzero $m$ the gaugino mass vanishes.  There is
a diagram in the low-energy theory which exactly cancels the contribution in the $m=0$ theory. Therefore, in interesting models of messengers, K\"ahler potential operators of the type considered in this subsection do
not generate leading-order gaugino masses.

\subsec{Spurion Models Beyond Leading Order}

In this subsection we will display the full, all-orders gaugino mass and
sfermion mass-squared for the weakly-coupled spurion models that
we considered in section 4 and the previous subsections of this appendix. The
full theory is defined by
\eqn\messthy{{\scr L} = \int d^4 \theta \, \Phi^\dagger_i(\delta_{ij} + V \tilde\lambda_{ij}) \Phi_j + \t \Phi^\dagger_i(\delta_{ij} + V \tilde\lambda_{ij}) \t \Phi_j + \int d^2 \theta \, (X \lambda_{ij} + m_{ij})  \Phi_i \t \Phi_j + {\rm c.c.}~.}
We will directly work in a basis in which $m$ is diagonal with real eigenvalues $m_i$. In this basis messenger parity and CP conservation require that $\lambda$ be real and symmetric; $\tilde\lambda$ must always be Hermitian. The $F$- and $D$-term spurions $X$ and $V$ acquire expectation values $X|_{\thetasq} = f$ and  $V|_{\theta^4} = D$ respectively, which we take to be real without loss of generality.

This model contains $N$ Dirac fermion pairs $\psi_i, \t\psi_i$ with masses $m_i$, and $2N$ complex scalars $(\phi_i, \t\phi^*_i)$ with mass matrix
\eqn\mmat{\CM = \pmatrix{m_i^2- D \tilde\lambda & - f \lambda \cr
                    -f \lambda & m_i^2- D \tilde\lambda}~.}
By a unitary transformation we can bring $\CM$ to block-diagonal form:
\eqn\newmmat{\CM \rightarrow \pmatrix{m_i^2 + f\lambda - D \tilde\lambda & 0 \cr
0 & m_i^2 - f \lambda - D \tilde\lambda} \equiv \pmatrix{\CM_+ & 0 \cr 0 & \CM_-}~.}
Since the matrices $\CM_\pm$ are Hermitian, there are unitary matrices  $U_\pm$ such that $U^\dagger_\pm \CM_\pm U_\pm = {\rm diag}(m^2_{\pm 1}, m^2_{\pm 2}, \ldots, m^2_{\pm N})$. To avoid tachyonic messengers, we need to ensure that $m_{\pm i}^2 >0$. In practice, this means choosing $f$ and $D$ sufficiently small compared to the supersymmetric messenger masses $m_i$. The supertrace is given by
\eqn\zerostr{\Tr \CM_+ + \Tr \CM_- - 2\Tr m_i^2 = -2D\Tr \tilde\lambda~,}
and we need to assume $\Tr \tilde\lambda = 0$ to render the model calculable.

Using the formulas in appendix B of \BuicanWS, we immediately obtain for the gaugino mass:
\eqn\mgaugino{M_{\t g} = - {g^2 \over 8 \pi^2} \sum_\pm \sum_{i,j = 1}^N (\pm) (U^\dagger_\pm)_{ij} (U_\pm)_{ji} \, m_j \, {m_{\pm i}^2 \log{m_{\pm i}^2 \over m_j^2} \over m_{\pm i}^2 - m_j^2}~.}

To calculate the sfermion mass-squared, we first compute the GGM
correlation functions defined in \Cfunsdef. From (B.4) in \BuicanWS\ we get:
\eqn\Cfunsii{\eqalign{\t C_0(p) & = \sum_\pm \sum_{i,j=1}^N (U^\dagger_\pm U_\mp)_{ij} (U^\dagger_\mp U_\pm)_{ji} I(p, m_{\pm i}, m_{\mp j})~,\cr
\t C_{1/2} (p) & = {1 \over p^2} \sum_\pm \sum_{i = 1}^{N} \Big(J(m_{\pm i}) - J(m_i)\Big)  \cr &\hskip45pt + {1\over p^2} \sum_{\pm} \sum_{i, j = 1}^N (U^\dagger_\pm)_{ij} (U_\pm)_{ji} (p^2 + m_{\pm i}^2 - m_j^2) I(p, m_{\pm i}, m_j)~, \cr
 \t C_1 (p) &= {1 \over 3 p^2}\Bigg( \sum_{\pm} \sum_{i = 1}^{N} \Big((p^2 + 4 m^2_{\pm i}) I(p, m_{\pm i}, m_{\pm i}) + 4 J(m_{\pm i}) \Big)  \cr & \hskip92pt + 4\sum_{i = 1}^N \Big( (p^2 - 2m_i^2) I(p, m_i, m_i) - 2 J(m_i) \Big) \Bigg) ~.}}
The Euclidian loop integrals $I(p, m_1, m_2)$ and $J(m)$ have been defined in (B.5) of \BuicanWS\ as follows:
\eqn\loopints{\eqalign{I(p,m_1, m_2) & = \int {d^4 q \over (2\pi)^4}\,  {1 \over ((p+q)^2 + m_1^2)(q^2 + m_2^2)}~, \cr
J(m) & = \int {d^4 q \over (2\pi)^4} \, {1 \over q^2 + m^2}~,}}
where a sharp momentum cutoff $q^2 \leq \Lambda^2$ has been imposed.
To obtain the sfermion mass-squared \ggmsoftmasses, we substitute the $\t C_a(p)$ of
\Cfunsii\ into \sfmass\ and do the momentum integral. Note that as $p^2
\rightarrow \infty$, the functions $\t C_a(p)$ in \Cfunsii\ only differ at
$\CO(1/p^4)$, so that this integral is guaranteed to be UV-convergent.\foot{This
is a general result which holds in any renormalizable theory: the difference of
any two $\t C_a(p)$ vanishes at least as rapidly as $1/p^4$ as $p^2 \rightarrow
\infty$. To prove this, we act on components of the current multiplet $\CJ$ in
\currsf\ with the supercharges to obtain the following two relations:
\eqn\corri{\sigmabar_\mu^{\alphadot\alpha}\vev{Q_\alpha \b Q_{\alphadot} \left(j^\mu(x)J(0)\right)} = 6 \d^2\left(C_0(x) - 2 C_{1/2}(x) + C_1(x) \right)~,}
\eqn\corrii{\vev{ Q_\alpha \b Q_{\alphadot}\left(j^\alpha(x)\b j^\alphadot(0)\right)} = -2 \d^2\left(C_0(x) + 2 C_{1/2}(x) - 3 C_1(x)\right)~.}
Consider the OPE of $j^\mu(x)J(0)$ as $x^\mu \rightarrow 0$. Since the current superfield $\CJ$ has dimension~$2$, we have
\eqn\OPEi{j^\mu(x)J(0) \sim {\CO x^\mu \over x^{-\Delta_\CO + 6}} + {V^\mu \over x^{-\Delta_V + 5}}+ \cdots~, }
where $\CO$ and $V^\mu$ are scalar and vector operators of dimension $\Delta_\CO$ and $\Delta_V$ respectively, and the dots denote less singular terms. By Lorentz invariance, only $V^\mu$ can contribute to~\corri, and moreover it cannot be a descendent. Hence $V^\mu$ must be a primary. The unitarity bound for a primary vector operator is $\Delta_V \geq 3$, which is saturated by a conserved current. Fourier transforming \corri, we conclude that the combination of the $\t C_a(p)$ on the right-hand side vanishes at least as rapidly as $1/p^4$ as $p^2 \rightarrow \infty$. The argument is completely analogous for the OPE of $j_\alpha(x) \b j_\alphadot(0)$, which contains a potentially different primary vector operator $V'^\mu$ with $\Delta_{V'} \geq 3$. This allows us to conclude that the result also holds for the difference of any two $\t C_a(p)$.
} To compute the relevant two-loop integrals we follow Martin \MartinZB. The
final answer is completely finite and can be simplified by using various
properties of dilogarithms. After the dust settles, we obtain for the sfermion
mass-squared:
\eqn\Afinal{\eqalign{& m^2_{\t f}  = {g^4 Y^2 \over 64 \pi^4} \sum_\pm \Bigg( \sum_{i = 1}^N \bigg( m^2_{\pm i} \log m^2_{\pm i} - m^2_i \log m^2_i \bigg) \cr & + \sum_{i, j = 1}^N \bigg( \half (U^\dagger_\pm U_\mp)_{ij} (U^\dagger_\mp U_\pm)_{ji} m^2_{\pm i} {\rm Li}_2 \Big(1 - {m^2_{\mp j} \over m^2_{\pm i}}\Big) - 2 (U^\dagger_\pm)_{ij} (U_\pm)_{ji} m^2_{\pm i} {\rm Li}_2 \Big(1 - {m^2_j \over m^2_{\pm i}} \Big)  \bigg) \Bigg) .}}
It is straightforward to expand these expressions to leading order in $D$ and
$f$, in which case they exactly reduce to the formulas derived in section 4 and subsections B.1, B.2. 

Formulas for the all-order soft masses in these models were first obtained in \MarquesYU, and we find complete agreement with \mgaugino\ and \Afinal. Special cases of these formulas have been
considered in \refs{\MartinZB,\NakayamaCF}.

\listrefs

\end